# Spectrum and Anisotropy of Cosmic Rays at TeV-PeV-energies and  Contribution of Nearby Sources

L.G. Sveshnikova[1] (corresponding author, tfl10@mail.ru,
+74959391391,+74959393553-fax),
O. N. Strelnikova[1],(olgov@mail.ru)

[1] *Skobeltsyn Institute of Nuclear Physics, Moscow State University, Leninskie Gory, 1, Moscow, 119192, Russia*

V.S. Ptuskin[2]

[2] *Pushkov Institute of Terrestrial Magnetism, Ionosphere and Radio Wave Propogation (IZMIRAN), Russian Academy of Sciences, Troitsk, Moscow region 142092, Russia,*
vptuskin@izmiran.ru

## Abstract

The role of nearby galactic sources, the supernova remnants, in formation of observed energy spectrum and large-scale anisotropy of high-energy cosmic rays is studied. The list of these sources is made up based on radio, X-ray and gamma-ray catalogues. The distant sources are treated statistically as ensemble of sources with random positions and ages. The source spectra are defined based on the modern theory of cosmic ray acceleration in supernova remnants while the propagation of cosmic rays in the interstellar medium is described in the frameworks of galactic diffusion model. Calculations of dipole component of anisotropy are made to reproduce the experimental procedure of "two-dimensional" anisotropy measurements. The energy dependence of particle escape time in the process of acceleration in supernova remnants and the arm structure of sources defining the significant features of anisotropy are also taken into account.  The essential new trait of the model is a decreasing number of core collapse SNRs being able to accelerate cosmic rays up to  the given energy, that leads to steeper total cosmic ray source spectrum  in comparison with the individual source spectrum.  We explained simultaneously the new cosmic ray data on the fine structure of all particle spectrum around the knee  and the amplitude and direction of the dipole component of anisotropy  in the wide energy range  1 TeV÷1 EeV.  Suggested assumptions do not look exotic, and they confirm the modern understanding of cosmic ray origin.

## Introduction

According to the standard model of cosmic ray (CR) origin [1], protons, nuclei and electrons are accelerated during the diffusive shock acceleration process by supernova blast waves. An increasing number of indirect indications that have appeared in recent years confirm that supernova remnants (SNR) are the main candidates for cosmic ray sources: the forward shock effectively accelerates cosmic rays [2],[3],[4],[5]. More than 100 galactic sources of TeV-gamma emission have been detected to date (http://tevcat.uchicago.edu ); most  are SNRs and pulsar wind nebulae (PWN) [3]. In many cases it is not yet clear whether electrons or protons emit such energetic gamma rays, and what the limit of cosmic ray acceleration in the Galaxy is. Nevertheless, the results of gamma-ray astronomy are in overall agreement with models of cosmic ray acceleration in SNR shocks [3], [4], [5].

Energetic particles accelerated in SNR leave remnants and diffuse in the interstellar magnetic fields filling the entire Galaxy. During the journey from the cosmic ray source to the Earth, cosmic rays are scattered on random galactic magnetic fields, and this makes CRs' direction distribution isotropic [1]. The time of propagation from a source to the Earth at low energy is much higher than the time for light propagation; therefore we have no chance to see directly cosmic ray sources in gamma rays as shells disappear in the interstellar medium at the age of tens of thousands years. Nearby sources can manifest themselves in anisotropy of cosmic rays caused by the gradient of cosmic rays. With increasing energy the propagation time decreases and at energy $10^{17}$-$10^{18}$ эВ the propagation speed becomes comparable with the speed of light and we can see the nearest cosmic-ray sources in gamma rays. At high energies the contribution of nearby sources can be seen in irregularities of cosmic ray fluxes [1]. The fluctuations in cosmic ray fluxes and anisotropy have been studied in many earlier papers [6], [7], [8], [9], and also more recently in [10] and [11]. It has been shown that fluctuations are caused mainly by the contributions of the nearest and youngest sources. If we account for their contribution we can calculate more accurately the prediction of anisotropies and fluxes.

An experimental base needed for this task was essentially enriched in recent years: new precise experimental data on CR characteristics (those shown to be very sensitive to the contributions of the local nearest cosmic ray sources) has appeared. Advanced, 'two – dimensional', cosmic ray anisotropy measurements were made in the northern sky in the Super-Kamiokande [12], Tibet [13], Milagro [14], and Argo-JBY [15] experiments; for the first time the Ice Cube collaboration [16], [17] presented anisotropy in the southern hemisphere in a wide energy range from multi-TeV up to 400 TeV. The anisotropy observed at 400 TeV shows substantial differences compared to those observed at a lower energy range. In addition to the large-scale features observed at 20 TeV in the form of strong dipole and quadrupole components, the data showed several localized regions of excess and deficit on scales between $10°$ to $30°$ [14]-[17]. Large-scale dipole anisotropy can provide unique information on the sources of CRs and the environment in which they have propagated, while the stochastic interstellar magnetic field probably plays an important role in the creation of median and small scale anisotropy [18] . The second important experimental achievement in cosmic ray measurements is an obvious fine structure of all particle spectra above the knee measured in the Tunka-133 [19],[20], KASCADE Grande [21],[22], , and Gamma [23] experiments. It consists of a 'hardening' around $2\times10^{16}$ eV and the subsequent 'second knee' (or even the bump-like structure [23]) around $10^{17}$ eV. One of the possible interpretations of this structure is a model where a 'single source' determines the shape of the knee [24], [25]. There is, however, no convincing indication of which source is responsible.

The goal of this paper is to calculate the spectrum and anisotropy of cosmic rays in the range $10^{12}$-$10^{18}$ eV, taking into account nearby SNRs from the latest gamma–ray astronomy catalogues, and to establish the limits of the contributions of some sources to the all particle spectrum in TeV to PeV energy range, including the knee region. We take into account the experimental procedure of 'two dimensional' anisotropy measurements and analyze, not only amplitude, but also the direction of anisotropy, and include in the consideration the time dependent emission of cosmic rays from supernova remnants that are very important for this task. We also take into account the arm structure of the Galaxy that is also very important for the investigation of the anisotropy and study a probable contribution of nearby sources around the knee together with the anisotropy analysis. The approximation of isotropic diffusion is used in our calculation that can be well justified only at scales more than 30 – 100 pc where locally anisotropic cosmic ray diffusion is isotropised by the action of a large – scale random magnetic field.

## 1. Method of calculation

### 1.1 Semi – statistical approach

The propagation of CR in the Galaxy is usually described in diffusion approximation in the model with a flat extended halo and continuous source distribution within the thin galactic disk [1], [9],[10],[11],[26]. A numerical realization of this model is the GALPROP code. This model explains the balance of cosmic ray energy in the Galaxy and the energy spectra of different species in a wide energy range [28]. To account for the contribution of actual nearby sources in this work a semi-statistical approach was applied [9]. All sources were divided into two groups: 'local' and 'background' sources. The first group was compiled from sources located within a distance of $R_{near}$ around the Earth and with an age of less than $T_{near}$. Sources were selected from different astronomy catalogues (see chapter 2); the second group was compiled from sources that were older than $T_{near}$ and located farther away than $R_{near}$. Their distances and ages were simulated randomly. The values $R_{near}$ ~1.5 kpc and $T_{near}$ ~ $5 \times 10^5 \div 10^5$ yr were chosen to provide a complete sample, because many more old and distant sources are missing and at present their shells are dispersed in the interstellar medium.

In our calculations we apply the Green function formalism that allows calculation the density and anisotropy of cosmic rays and their fluctuations as a sum of contributions of different individual sources [1]. We use a simple flat-halo galaxy model of cosmic ray transport with halo boundaries at the height $|H_z| = 4$ kpc and the galactic radius $R_G = 15$ kpc, where cosmic rays freely exit from the galaxy. The cosmic ray proton number density $N(E, t, \boldsymbol{r})$ obeys the diffusion equation [1].

$$\frac{\partial N}{\partial t} - \nabla(D\nabla D) = Q(E, t, \boldsymbol{r}) \qquad (1)$$

Here $D$ is the cosmic ray diffusion coefficient and $Q$ is the source term. While the consideration concerns only a high-energy region, we ignore energy losses, advection and secondary nuclei contribution as well as reacceleration processes. The cosmic ray density $N$ and the cosmic ray flux $F = (c/4\pi) N$ near the Earth are calculated as a sum of Green functions $G(t_i, r_i, E)$ that are calculated in Appendix A:

$$F = \frac{c}{4\pi} \sum_{i=1}^{Nsour} G_i(t_i, r_i, E) \qquad (2)$$

$$G_i(t_i, r_i, E) = Q_i(E, E_{max}) \frac{\exp\left\{-\frac{r_i^2}{4R_{dif}^2}\right\}}{4\pi H R_{dif}^2} \times S_i \qquad (3)$$

$$S_i = \sum_{n=0}^{\infty} \exp\left(-\frac{(2n-1)^2 \pi^2 R_{dif}^2}{4H^2}\right) \qquad (4)$$

$$Q(E, E_{max}) = Q_o E^{-\gamma} \left(1 + \frac{E}{E_{max}}^{\omega}\right)^{-d\gamma/\omega}$$

$$Q_0 = (\gamma - 2) k_A W_{sn} m_A^{(\gamma-2)} \qquad (5)$$

Here the i-source (from the total number of $N_{sour}$) is located at the distance $r_i$ from the Earth and instantly emits cosmic rays at time $t_i$ with the spectrum $Q(E, E_{max})$. The total energy of emitted relativistic particles $\int dt \int_{Am_p^2} dE \, E \, N_1(E, t) = k_A W_{sn}$, where $W_{sn}$ (=$10^{51}$ erg) – kinetic explosion energy; $k_A$ - the part of energy converted to the cosmic ray nuclei with mass $m_A = A m_p$ ($m_p$ is the proton mass); and charge Z (for protons $k_p \approx 0.15$). $E_{max}$ - is the maximum energy of accelerated particles in the source spectrum $Q$; $\gamma = 2.2$ is the slope of the spectrum up to the $E_{max}$; $\gamma + d\gamma$ is the slope above the $E_{max}$ energy; $\omega$ denotes the 'sharpness' of this change. To interpret the pronounced structures around the knee, such as the first knee at 4 PeV and a hardening at 20 PeV, we need to have a very sharp cut-off in the source spectrum and we need a parameter in our approximation that allows to change the sharpness of the cut-off. The parameter ω was introduced for this purpose. The widely used a pure exponential cut-off would lead to smoother change at $E_{max}$. Besides, when one takes into account the process of particle escape (see discussion in the next Section) the overall spectrum in general does not have an exponential cut-off. This spectrum is the superposition of particles advected downstream and released later plus the particles escaping from upstream region [29]. In (5) the slope is changed from 2.2 to 4.2 at energy $E_{max}$, and $\omega \geq 4$. This sharp non exponential cut-off does not contradict the source spectrum calculated in [30].

The theory of nonlinear shock acceleration predicts the concave particle spectrum with the slope < 2.0 at very high energies for strong hydrodynamic shocks. However, the gamma-ray observations of young supernova remnants and the empirical model of cosmic ray origin require more steep spectra, typically 2.2 … 2.4. The theory requires complication to explain more steep instant gamma-rays spectrum measured in the Tycho's remnant [86], [87]. Steep spectra of this kind can be theoretically obtained when the effects of strong magnetic field amplification by cosmic ray streaming instability near the shocks and some other factors are taken into account, see e.g. [27] for discussion. The spectrum with a slope close to 2.2 (although it did not have a pure power law shape) was found in the modeling of cosmic ray acceleration [28] where the Alfvenic drift of particles upstream and downstream of the shock was taken into account. Our present consideration is based on this result.

Factor $S_i$ takes into account the halo boundaries (at $H << R_G$) (see Appendix A). The diffusion radius

$$R_{dif} = \sqrt{D(E)_Z t_i} \qquad (6)$$

depends on the energy through the diffusion coefficient, $D_Z(E) = D_p(E/Z)$. We have chosen the energy dependence corresponding to the energy dependence of a diffusion coefficient in the 'model with reacceleration' [26] at energy $E>30$ GeV; that is, $D(E)$ has a power like form

$$D_p(E) = 3.3 \times 10^{28} \, E^{0.33} \quad \text{cm}^2/\text{s} \qquad (7)$$

The reacceleration of cosmic rays is insignificant at the high energies under consideration. The approach used here requires assumptions about the birthrate of CR sources and the time of cosmic ray confinement in the galaxy. We chose $T_{conf}= 1.4 \times 10^8$ years at 1 GeV/nucleon, which corresponds to the Halo boundaries $H_z=4$ kpc for the diffusion coefficient (7) and does not contradict the confinement time obtained from the abundance of secondary decaying isotopes in cosmic rays [1]. Assuming the SN rate in the galactic disk to be 1/50 yr over the entire Galaxy, we simulated $3 \times 10^6$ sources distributed in space and time in one random realization. For the study of fluctuations we usually simulated 10-20 realizations of source distributions and added the fixed sample of nearby and young sources from the Table 1.

### 1.2. Energy dependence of particle escape time

In the theory of cosmic ray diffusive acceleration, a cosmic ray streaming instability can cause significant amplification of a turbulent magnetic field ahead of the moving shock [31], [32], [33], [27], [29] leading to an increase in the maximum energy of accelerated particles, $E_{max}$, above the Bohm limit at high shock velocity. At that time the maximum energy of accelerated particles decreases with SNR age, when the shock velocity and the cosmic ray induced turbulent magnetic field decrease due to the magnetic field damping [33]. At the given moment of remnant evolution, $T_{esc}=T_{SNR}$, the particles with energy larger than $E_{esc}$ ($T_{esc}$) leave the system from upstream region (the escaping particles). They carry away a large fraction of energy if the shock is strongly modified by the presence of cosmic rays [29],[33]. The remaining particles are advected downstream and undergo adiabatic energy losses before being injected into the ISM [29]. The flux of escaping particles plays an essential role in the formation of the cosmic ray spectrum detected at Earth, see [30]. The cosmic rays trapped in supernova shells experience adiabatic energy losses and would probably have energies below the knee after release from the breaking up SNRs. The nonlinear and linear wave damping may [33] considerably suppress the level of turbulence at $U < 10^3$ km s$^{-1}$ and thus may decrease the maximum particle energy much below the Bohm limit. As a result, $p_{max}(U)$ dependence has a non power law form. In Figure 1 (top panel), we present the dependence of $T_{esc}(E_{esc})$ obtained from the calculations of $p_{max}(U)$ and $U(T_{SNR})$ for SNR Ia in [32]. $T_{esc}(E_{esc})$ also has a non power law form and can be approximated by the following polynomial function:

$$\log T_{esc} = 5.47 - 0.29 \log E_{esc} + 0.047 \log E_{esc}{}^2 - 0.02 \log E_{esc}{}^3 + 0.001 \log E_{esc}{}^4. \quad (8)$$

The instant spectrum of escaped particles at the moment $T_{esc}$ was calculated as a delta-function $\delta(E-E_{esc}(T_{esc}))$. The dependence $E_{esc}(T_{esc})$ is not well known; we considered also a simple power law form for it (as in [10])

$$T_{esc} = T_{max}\left(\frac{E_{esc}}{E_{max}}\right)^{-\frac{1}{\alpha}}. \quad (9)$$

In (9) $E_{max}$ is a maximal energy reachable in the remnant at the beginning of the Sedov stage, at $T_{max}$. For example, at $E_{max} \sim 4\times Z$ PeV and $T_{max}=100$ year and, on condition that particles with energy $40\times Z$ GeV escape from the remnant at $t = 3\times 10^4$ years, the slope index $\alpha = 2$.

The Green function can be written now as Eq. (3), but $t_i$ is replaced by $\tau_i(E) = t_i - T_{esc}(E)$ and the diffusion radius (6) is replaced by $R_{dif}=(D(E)\cdot\tau(E)_i)^{1/2}$.

The bottom panel in Fig. 1 presents for illustration the spectra of protons near the Earth produced by two nearby sources, important for our consideration: Vela X located at 290 pc and with the age 11 kyr, and Vela Jr. if it is located at 0.3 kpc and has an age 0.7 kyr. Three cases are shown: instant emission of particles; time dependent emission (8); and the power law time-dependent emission (9) with, $E_{max}=4$ PeV, $\alpha=2$. It can be seen that, in both cases of energy dependent escape time, the spectra are depleted in low energy particles that are confined in the remnant. Vela Jr. can produce (Fig. 1) some 'bump' around the knee, because particles with energy less than 100 TeV are confined in the shell; the expected flux is comparable with the total flux measured around the knee. Vela Jr. therefore might be a good candidate for the knee producer if it is so young and nearby. This possibility will be discussed in the third chapter.

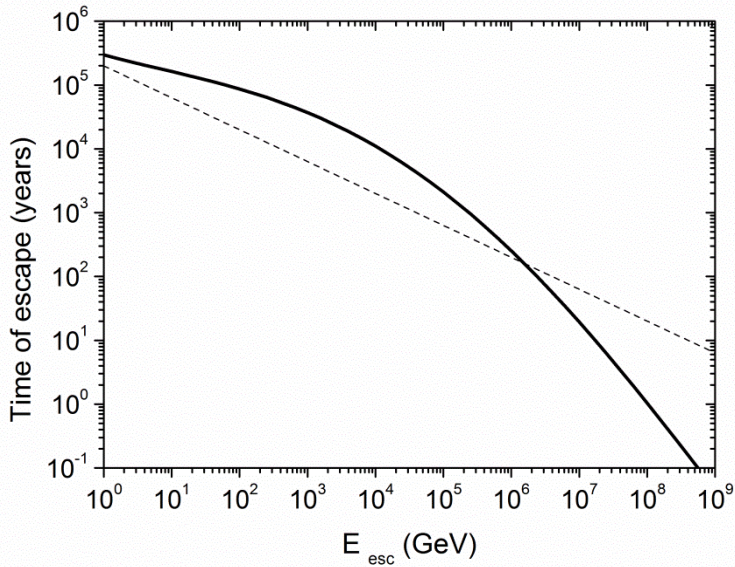

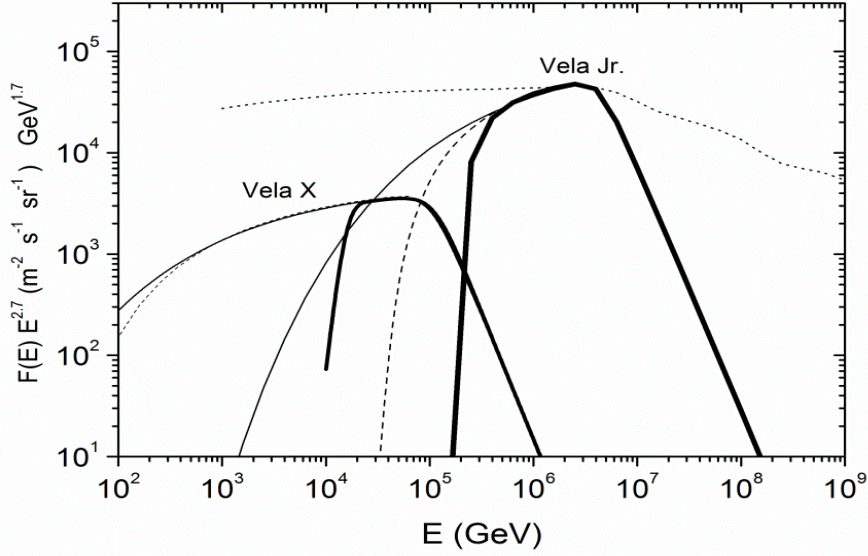

**Fig. 1.** [Top panel]: $T_{esc}(E_{esc})$ dependence (8) (thick line) and the power law dependence (9) for $\alpha=2$, $E_{max}=4$ PeV, $T_{max}=100$ yr (dashed line). [Bottom panel]: spectra of protons near the Earth produced by Vela X ($R$=0.29 kp, age=11 ky) and Vela Jr (if $R$=0.3 kpc, age=0.7 ky) for 3 cases: instant release of all particles (thin lines); $T_{esc}(E_{esc})$ (8) (thick lines); the power law dependence (9) with $\alpha=2$ (dashed lines). The dotted line denotes the approximation of the experimental all particle spectrum from Fig. 4.

The flux from the remnant Vela X can also reach several percents of the total flux and has noticeable influence on the anisotropy at multi-TeV range (see discussion in chapter 4).

From Fig. 1 (bottom panel) it is seen that a non power law dependence on energy of the escape time used in our work gives more pronounced 'bumps' produced by nearby sources in comparison with the pure power law dependence proposed in the recent work [10].

### 1.3. Can we see cosmic ray sources?

Gamma-ray astronomy, in principle, provides a direct view of the astrophysical accelerators of cosmic ray nuclei, because they can interact with the surrounding matter and produce secondary gamma rays via $\pi^0$ decay [2]. But the times of observation for gamma-rays, propagating with the speed of light, and cosmic rays, propagating diffusively and much more slowly, are shifted significantly. Figure 2 presents the density of protons near the Earth, produced in one SNR located at $R$=1 kpc and observed at the Earth at the time $T$ after the SNR explosion (the time dependent emission is also taken into account). It can be seen that the propagation time from the distance 1 kpc to the Earth is less than $10^5$ years for particles with multi-TeV energies; this is comparable with the life time of SNR shells. This means that multi-TeV cosmic ray sources and gamma ray sources can be observed simultaneously within the radius 1 kpc. The sources of particles for PeV energies can be seen within the radius 1.5 kpc, because the propagation time is much less than a life time of SNR shells. The efficiency of detection for gamma sources located farther than 1.5 kpc falls rapidly with distance. Based on these estimates, we selected the candidate CR sources from the gamma-ray catalogues.

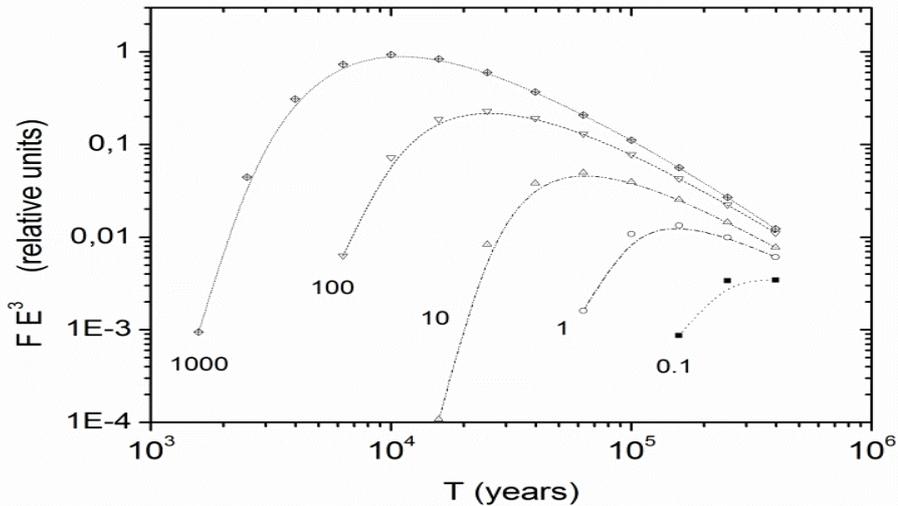

**Fig. 2.** Density of protons with different energies E (in TeV) near the Earth at the time $T$ after SN explosion, multiplied by $E^3$ (in relative units).

## 2. Candidate sources of cosmic rays

### 2.1. Types of sources

In preparing the catalogue of nearest potential CR sources we used the conventional classification of SNe [5], [34], [35], [36], and grouped sources into two different types in accordance with their physical nature: core collapse SNe; and thermonuclear explosions of accreting white dwarfs after their mass approaches to the Chandrasechar limit ~ 1.4 $M_\odot$ (SNeIa). The SNRIa type can be seen in a wide range of electromagnetic wavelengths (from radio to TeV gamma rays as 'shell' SNR); their energies of explosions are $10^{51}$ erg. They can travel far away from the star formation zone due to the long life of their progenitors (white dwarfs) and can be observed between galactic arms [37]. They constitute about 20-30% of all SNe [5], [34], [35]. The SNRIa can be searched in the Green's radio catalogue of galactic supernova as 'shell' SNR [38], [39]. The historical supernovae in our Galaxy, such as SN1006, Tycho and Kepler, are of this type.

Core collapse supernovae, SNIb/c, SNII (L,P, n, b subtypes), arise in gravitational collapses of massive stars ($M > 8\ M_\odot$) leaving a compact remnant, that is a neutron star (in most cases a magnetized rotational pulsar), or a black hole. They comprise the bulk of exploding stars in the Universe (75-80%) and demonstrate a great variety of their properties. [34], [35], [36]. There are several hypernovae detected in other galaxies with explosion energies up to $60 \times 10^{51}$ erg [36]. The core collapse SNe can be searched for as 'composite' or 'mixed morphology' SNe in the Green's catalogue [5], [38].

At the late stage of evolution ($T > 10^5$ years), when the envelope has already dispersed in the ISM, a core collapse SNe can be seen only as a pulsar. The number of known radio pulsars is more than 1800 in the ATNF catalogue [40]. There is another group of pulsars, detected at X-ray and gamma-ray wavelengths. A sharp increase in the number of known gamma-ray pulsars has occurred in recent years due to the Fermi Gamma-ray Space Telescope observations (about 46 gamma pulsars) [41].

A very important class of potential sources is pulsar wind nebulae (PWN) [42]. They are produced, in some cases, by young power pulsars and are observable from radio through gamma-ray bands of waves. All active pulsars lose their spin energy and angular momentum via relativistic winds composed of relativistic particles and electromagnetic fields. Since a pulsar is formed during an SN explosion, the star and its PWN are initially surrounded by an expanding SNR, sometimes seen as a shell-like SNR (in this case the system is referred to as 'composite'). In other cases, best typified by the Crab itself, no surrounding shell is seen. The PWN catalogue of 46 PWN detected by Chandra was used [42].

The most relevant catalogues for our task are those of TeV sources, because they provide a direct view of the astrophysical accelerators of charged particles and allow the identification of individual sources of cosmic rays. The Imaging Air Cherenkov Telescopes, H.E.S.S [2], [43], MAGIC [44], and VERITAS [45] have confirmed the anticipated detection of a few supernova remnants and have established a rich and diverse collection of VHE sources. It is now clear, however, that only very young shell SNR can be detected in the TeV range. A possible reason for this is the relatively short time of TeV emission [32], requiring a large shock velocity. In preparing the combined catalogue of potential CR sources we used Green's radio catalogue of 274 SNRs [38], including 174 SNRs with measured distances [39]; the X-ray catalogue of 54 PWNe [42]; the Fermi-LAT catalogue of 46 gamma-pulsars [41]; the ATNF of 1827 radio pulsars [40]; the catalogue of TeV sources of HESS [43]; and some new TeV sources, detected in MAGIC [44] and VERITAS [45] experiments. The ages and distances of sources with pulsars were determined more or less reliably by pulsars and we accepted these parameters from the catalogues [40], [41], [42]. Regarding the pure shell SNRs, however, distances and ages were determined with large uncertainties [38], [39]. The total number of sources with estimated ages and distances was 73. Sources located further than 1.5 kpc are detected very inefficiently and were not included in our list. In Table 1 the minimal and maximal estimates of the distances are shown, but in calculations we used only minimal estimates to derive the maximal possible contributions of nearby sources.

The nearby sources with $R_{near}$ <1.5 kpc and $T_{near}$ <$10^5$ years are listed in Table 1. The following information is shown for every source: name of primary catalogue (most of the sources are seen in several catalogues); age; minimum and maximum distances estimated in different ways; the source's coordinates in the galactic coordinate system, and in the second equatorial system (right ascension, $\alpha$, and declination, $\delta$). In the penultimate column we also present an approximate estimate of the energy $E_{mc}$, based on the Eq. (3-8), at which the density of cosmic rays generated in this source reaches a maximum near the Earth at the present time.

It is interesting to note that for many of the sources listed below the energy, at which the density of cosmic rays generated in the source approaches maximum near the Earth, is close to the knee region, and so they may be studied as possible knee producers.

**Table 1. The list of selected potential CR sources**. Columns: 1-number in the list, 2 – catalogue, where this source can be found ("S" – Green's SNR catalogue [38], [39], "W"- PWN catalogue [42], "H"- TeV emission [42] - [45] and others, "F"- Fermi-Lat gamma-pulsars catalogue [41], "P"-ATNF catalogue [40]), 3 – galactic longitude, 4 - galactic latitude, 5 – $\alpha$, right ascension, 6 - $\delta$, declination, 7- minimal estimate of distance, 8- maximum estimate of distance, 9 – estimate of age, $E_{mc}$ - energy, at which the density of cosmic rays generated in this source, approaches maximum near the Earth location, 10 - name.

| N | HSWFP | L deg. | B Deg. | $\alpha$ hour | $\delta$ Deg. | $D_{mn}$ kpc | $D_{mx}$ kpc | T kyr | $E_{mc}$ TeV | Name |
|---|---|---|---|---|---|---|---|---|---|---|
| 1 | _SW__ | 65.3 | 5.7 | 19.6 | 31.3 | 0.8 | 1.5 | 20. | 400 | G65.3+5.7 |
| 2 | _SW__ | 65.7 | 1.2 | 19.9 | 29.4 | 1.5 | 1.5 | 60 | 800 | G65.7+1.2; DA495 |
| 3 | _S___ | 74.0 | -8.5 | 20.8 | 30.4 | 0.56 | 0.77 | 10. | 100 | Cygn Loop |
| 4 | HS_F_ | 78.2 | 2.10 | 20.4 | 40.4 | 1.5 | 2.0 | 7. | 316 | G78.2+2.1; DR4;Cigni |
| 5 | _S___ | 89.0 | 4.7 | 20.8 | 50.7 | 0.8 | 0.8 | 19. | 504 | G89.0+4.7; HB21 |
| 6 | HSWFP | 106.3 | 27 | 22.5 | 60.8 | 0.8 | 3.1 | 10. | 4000 | G106.6+2.9; Boomerang |
| 7 | _S__P | 114.3 | 0.3 | 23.6 | 61.9 | 0.7 | 2.47 | 7.7 | 3980 | G114.3+0.3 |

| 8 | HSWF_ | 119.5 | 10.2 | 0.1 | 72.8 | 1.4 | 1.4 | 14. | 39000 | G119.5+10.2; CTA_1 |
| 9 | _S___ | 127.1 | 0.5 | 1.5 | 63.1 | 1.15 | 1.2 | 30. | 1600 | G127.1+0.5; R5 |
| 10 | _S___ | 160.9 | 2.6 | 5.0 | 46.2 | 0.8 | 1.2 | 6.6 | 12600 | G160.9+2.6; HB9 |
| 11 | _SW_P | 180.0 | -1.7 | 5.7 | 28.0 | 0.8 | 1.5 | 40. | 63 | G180.0-1.7; S147 |
| 12 | HSW__ | 189.1 | 3.0 | 6.3 | 22.5 | 1.5 | 1.5 | **20**-30. | 20000 | IC443, 3C157 |
| 13 | HSWFP | 263.9 | -3.3 | 8.6 | -45.8 | 0.29 | 0.32 | 11.0 | 31 | G263.9-3.3; Vela X |
| 144 | HSWF_ | 266.2 | -1.2 | 8.9 | -46.3 | 1.0(0.3) | 1.3 | 10(0.7) | 1.58e6 (4000) | G266.2-1.2; Vela Jr. |
| 15 | HWFP | 343.1 | -2.3 | 17.1 | -44.3 | 1.4 | 2.0 | 18. | 20000 | Fermi-G343.1 |
| 16 | HS___ | 347.3 | -0.5 | 17.2 | -39.8 | 1.0 | 1.3 | 1.6 | 4e6 | J1713-3946 |
| 17 | ___P | 49.1 | 0.87 | 19.3 | 14.8 | 1.4 | 1.4 | 88. | 160 | PSRB1916 |
| 18 | ___FP | 201.2 | 0.5 | 6.5 | 10.7 | 0.75 | 3.62 | 44. | 39 | J0631+1036 |

**2.2. Statistics and birthrate of nearby young sources**

The birthrate of different types of sources is known with an accuracy of approximately 30%. According to [41], the gamma-ray-selected young pulsars are born at a rate comparable to the rate of radio-selected pulsars; the rate of all young gamma-ray-detected pulsars (~1/50 yr) is similar to that of galactic supernovae. Other estimates give the pulsar birthrate (0.9-1.9)/100 yr [46]. One of the latest estimates of SN rate from the Lick Observatory Supernova Search [35], [47] gives the following SN rate per century: 0.54±0.12 (SNIa), 0.76±0.16 (SN1bc), 1.54±0.32 (SNII), and 2.30±0.48 (core collapse SNe). The total SN rate is estimated as 2.84±0.60. The core collapse rate, 2.30±0.48, coincides very well with the estimates made in [41] (~1/50 yr).

Among 18 selected sources only six (numbers 3, 5, 9, 10, 14, 16 cited in Table 1) are likely to be related to SNRIa, because they have no pulsar or PWN. This upper limit does not contradict the expected ratio of SNIa among all SNe (20-30%) [34],[35]. The other 12 sources can be identified as remnants of core collapse SNe. TeV emission is observed only in the one shell SNR J1713-3946 and in nine core collapsed SNe.

To check the completeness of the sample of sources with distances $R$<1.5 kpc and ages $T$<$10^5$ yr, we compared the expected distance-age distribution with the experimental distribution and found that approximately 50% of sources with ages $6\times 10^4$ yr to $1\times 10^5$ yr are lost, probably because their shells have dispersed in the interstellar medium during this period. Table 1 shows there are 16 sources with ages $T$<6 $10^4$ yr. The expected total number of sources with a distance $R$<1.5 kpc and an age $T$<6 $10^4$ yr is 16±4 for the birthrate 1/50 yr. This number coincides very well with the experimental number.

**2.3. Spatial distribution of sources**

A spatial distribution of all SNRs probably has a maximum at 3–4 kpc [39]. The core collapsed SNe are more tightly concentrated to the arms than SNe Ia, as established in other spiral galaxies [48], [49]. SNe Ia are old enough to diffuse away from their formation regions. Positions of core collapse SNe were simulated inside the arms, while SNIa were simulated uniformly over the disk. Unfortunately, a spiral arm structure of the Galaxy is not well established. Based on data obtained via the Spitzer telescope [50] we introduced six arms: 'Outer'; 'Perseus'; 'Orion'; 'Sagittarius'; 'Scutum-Centaurus'; and 'Center'. The outer and the inner boundary of every i-arm is expressed by a spiral function $R_i(\varphi)$:

$$R_i(\varphi) = R0_i \times 10^\varepsilon\ ;\ \ \varepsilon = \beta_i(\varphi - 45)\ ;\qquad(10)$$

where $\varphi$ is the azimuthal angle in degrees counted from the $X$ axis; $R0_i$ is the boundary distance at $\varphi = 45°$ and $\beta_i$ are parameters. Changing these parameters, it is possible to change the radial distribution of sources. The Earth is located near the inner boundary of the Orion arm at $Y=8.5$ kpc ($X,Y$ – local coordinate system).

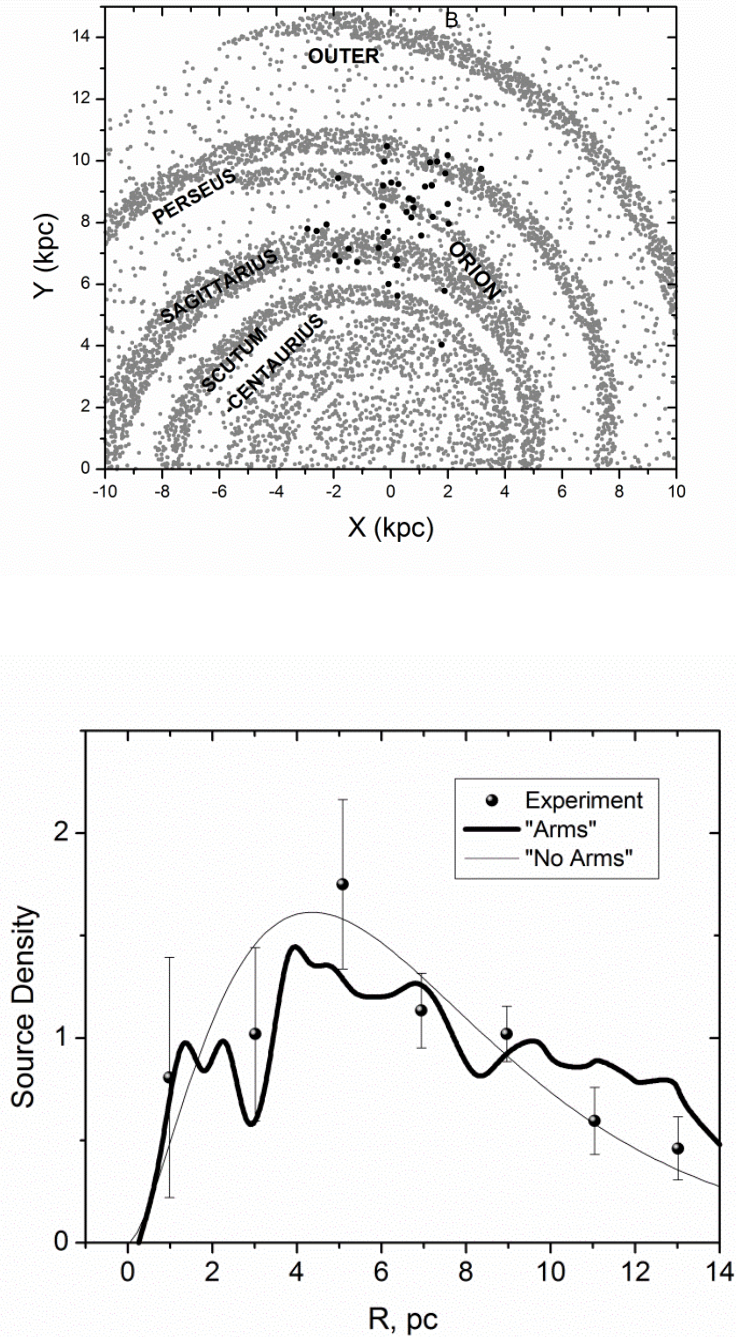

**Fig. 3**. [Top panel]: spatial 'arm' distribution of background sources (gray points) and of actual young sources with $T<10^5$ years and $R< 3$ kpc (black circles). [Bottom panel]: radial distribution of background sources 'No arms' (thin line) and 'Arms' (thick line); experimental SNR distribution from [39] (black circles).

In Fig. 3 (top panel) the two-dimensional $X$-$Y$ distribution of randomly simulated sources are shown; the bottom panel of Fig. 3 shows the radial distribution of sources for the two cases, named 'Arms' and 'No arms'. 'No arms' corresponds to the smooth distribution similar to the SNR distribution obtained in [39]. Actual young ($T<10^5$ years) sources from our list of sources with estimated distances of less than 3 kpc are shown as black circles. One can see that young actual sources are concentrated to the spiral arms.

### 2.4. $E_{max}$ in source spectrum in different SNRs

To analyze an anisotropy and irregularities of CR flux, it is necessary to take into account two additional factors: the number of sources able to accelerate particles up to given energy E, and the chemical composition, since fluctuations of CR depend on the number and density of sources (as $q^{-1/2}$) [1], [8], [9] and on particle charge Z via the diffusion coefficient $D\sim (E/Z)^{0.33}$. At energies of more than 100 TeV only EAS data are used for anisotropy measurements, and so a contribution of primary nuclei can be significant.

In our approach we use the source spectrum in one remnant in the form (5). The acceleration of CR is calculated in [28], [30] together with the evolution of a supernova blast wave in the framework of the numerical code that includes the hydrodynamic equations solved simultaneously with the diffusion–convection transport equation for the cosmic-ray distribution function. This model took account of a cosmic-ray streaming instability resulting in strong amplification of random magnetic fields in the vicinity of the shock and a significant increase of maximum energy of acceleration.

The maximum particle momentum depends on the shock velocity $U_{sh}$, the shock radius $R_{sh}$, the kinetic energy $E_{51}$ (normalized to $10^{51}$ erg), the gas number density $n$, and the mass of supernova ejecta, $M_{ej}$: $E_{max}/Z \sim 20(U_{sh}^2)R_{sh}n^{1/2}$ TeV, where the shock radius is measured in pc, the interstellar gas number density in cm$^{-3}$, $U_{sh}$ in $10^3$ km/s, $M_{ej}$ in solar masses $M_\odot$. For the shell SNRs exploding in the interstellar medium one has $E_{max} \sim E_{51}n^{1/6}M_{ej}^{2/3}$ PeV [28]. Ib/c is exploding into the low-density bubble with density $n = 0.01$ cm$^{-3}$ formed by a progenitor like a Wolf-Rayet star. As a result, the $E_{max}$ value differs for different types of SNRs (see Table 2).

**Table 2.** Maximum CR energy achievable in different types of SNe calculated in [28].

| SNR type | $E_{max}$ (PeV) | $M_{ej}$ ($M_\odot$) | n (cm$^{-3}$) | $E_{51}$($10^{51}$erg) |
|---|---|---|---|---|
| Ia | 4×Z | 1.4 | 0.1 | 1 |
| Ib/c | 1×Z | 2 | 0.01 | 1 |
| IIp | 0.1× Z | 8 | 0.1 | 1 |
| IIb | 300× Z | 1 | 0.01 | 3 |

A picture of cosmic ray production in SNRs used in the present work is based mainly on the paper [28] that considered four different types of supernova remnants (SNR IIP, SNR Ia, SNR Ib/c, and SNR IIb) with fixed parameters of explosions for each type of SN. The maximum energies $E_{max}$ of accelerated particles in the overall spectra of cosmic rays produced in these SNRs over their lifetimes are also fixed and differ by about 3 orders of magnitude. The spectra of cosmic rays observed at the Earth were reproduced with the dependence of diffusion coefficient on energy $D \propto E^{0.54}$ typical for the empirical plain diffusion model of cosmic ray propagation in the galaxy. This dependence is strong and probably not consistent with a small value of observed cosmic ray anisotropy, which is more compatible with the diffusion coefficient $D \propto E^{0.3}$ typical for the diffusion model with reacceleration, see [9], [10], [11]. The process of distributed interstellar reacceleration in itself should be taken into account in the cosmic ray transport equation at energies below about 100 GeV and is not important for our calculations.

To combine the scaling of diffusion $D \propto E^{0.3}$ with a hard source spectrum ($\gamma_{sour} = -2.2$) and observed cosmic ray spectrum ($\gamma \sim = -2.7$), we introduce the distribution of maximum energy of accelerated particles $E_{max}$ even within each type of core-collapse SNRs. The following observational properties of core collapse SNe may support this invention.

1) There is strong evidence [36] for a continuous spectroscopic sequence (II-IIb-Ib-Ib/c-Ic) among core collapse SNe, which reflects the ability of the SN progenitors to retain their H-rich and He-rich envelopes prior to explosion [36]. Another factor that defines the appearance of a SN is the density of the medium in which they explode, which is determined by the history of mass-loss of the SN progenitor [34], [36].
2) There is a wide range in explosion energies (0.6-60 $10^{51}$ erg), ejected masses, and $^{56}$Ni yields (0.0016-0.5 $M_{ej}$) among core collapse SNe, even within the same spectroscopic type [36]. But even for narrow II-P sample as in the work [34] initial shock velocity varies from 1500 km s$^{-1}$ up to 6000 km s$^{-1}$ that can lead to a big difference in $E_{max}$ value.
3) There is the class of SNe (~10-15%) where surrounding outer shell that was left after SNR explosions is not seen [51], [52], [53], [54]. From 309 galactic SNRs from catalog of X-ray and gamma-ray observations [52] about ~90 SNRs contain plerions or plerionic candidates, about 30 of which lack shells (including the Crab). The famous PWN Crab Nebula has no visible outer shell, that would appear to conflict with the high luminosity indicated by historical observations of SN 1054 produced this PWN [5], [51], [53], [54]. While the standard explanation for this fundamental puzzle is invisibility of the shell [55], many others hypothesis were suggested to explain this fact [51], [54], [56], including the important for our consideration hypothesis of sub-energetic explosions proposed recently in [54]: Crab Nebula is the end product of a Type IIP-n SN explosion, where the explosion mechanism was the collapse of a degenerate O-Ne-Mg core that yields a sub-energetic ($10^{50}$ erg) explosion. But their high visual-wavelength luminosity and Type IIn spectra are dominated by shock interaction with dense circumstellar material rather than the usual recombination photosphere of a SN.

Consequently, we have a reason to take into account the probable dispersion of $E_{max}$ values for cosmic rays accelerated in core collapse supernovae, due to a great variety and continuity of their parameters of explosions [34], [36].

As the first approximation, we introduce the distribution of core collapse SNRs on $E_{max}$ so that the fraction of SNRs that are able to accelerate cosmic ray nuclei up to maximum energies not larger than $E_{max}$ decreases with energy as $\sim E_{max}^{\delta}$. We are based on the next reference points: only SNRs of the type Ia can accelerate protons up to $E_{max}$=4 PeV (20%), SN Ib/c have $E_{max}$ distributed in the range from 100 TeV to 4 PeV (20%), SNRs of II type have $E_{max}$ distributed in the range from 1 to 100 TeV (55%), including ~18% of Crab-like SNRs, which can accelerate particles up to energies not higher than 3 TeV. The value $\delta \sim -0.17$ approximately fits these reference points up to the knee. The resultant spectrum observed at the Earth after propagation in the galaxy has a slope $\gamma_{obs}$ (up to the knee):

$$\gamma_{obs} = \gamma_{sour} + d\gamma_{prop} + \delta \cong -2.7.$$

Here $\gamma_{sour} = -2.2$, $d\gamma_{prop} = -0.33$ is a steepening caused by the propagation effects, the decreasing number of sources accelerating particles up to given $E_{max}$ results in an additional steepening by $\delta = -0.17$.

### 3. All particle spectrum

The all particle spectrum calculated with randomly distributed values of $E_{max}$ with parameter $\delta = -0.17$ is presented in Figure 4 in the top panel, where the contributions of different types of SNR are shown separately. In the bottom panel we present the corresponding average logarithmic mass that will be commented upon further. Fluctuations caused by the random nature of sources are less than 10% up to energy $10^{17}$ eV and will be discussed below in greater detail. For illustration, a few of the most recent experiments are shown: ATIC2 [57]; Tibet (QGSJET model) [58]; Tunka -25 [59]; Tunka-133 [19], [20]; and KASCADE Grande [21]. The calculated all particle spectrum fits these experiments well, including a sharp shape of the main knee, as well as the change of the all particle spectrum slope at $2\times10^{16}$ eV (from $\gamma = -3.2$ to $\gamma = -3.0$) and at $\sim10^{17}$ eV (from $\gamma = -3.0$ to $\gamma = -3.2$), as it was recently observed in the Tunka 133 [19], [20] and KASCADE Grande [21], [22], experiments. The fine structure at the knee region and above the knee is reproduced only if the relatively sharp cutoff in the source spectrum of SNIa is assumed: $\gamma = -2.2$, $d\gamma > 2$, $\omega > 4$, $E_{max} = 4\times Z$ PeV in Eq.(5). This suggestion does not contradict the theoretical predictions for the source spectrum obtained in [30].

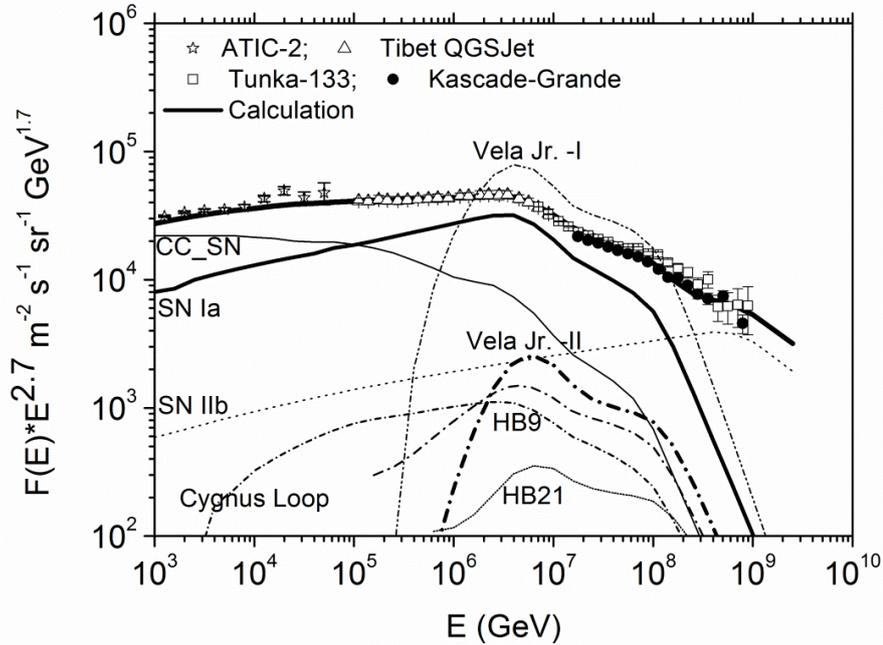

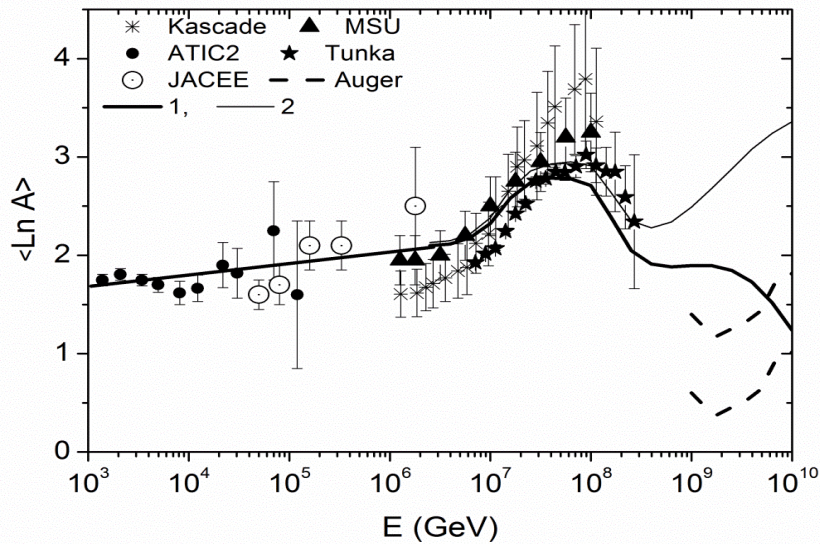

**Fig. 4.** [Top panel]: All particle spectrum as measured in ATIC2 [57], Tibet QGSJET [58], KASCADE-Grande [21], [22], and Tunka-133 [19], [20] experiments (symbols). Calculated all particle spectrum (very thick line) includes contributions of different SN types: SN Ia (thick line), CC SN (thin line), SN IIb (dotted line) and contributions of individual shell SNRs: Cygnus Loop, S147, HB9, HB21, Vela Jr. (-I, if R=0.3 kpc, T=700 yr; –II, if R=0.7 pc and T=1.7 yr ). [Bottom panel]: <LnA> (E) dependence corresponding to all particle spectrum for two cases: 1 (without extragalactic protons) and 2 (with extragalactic protons). Some experimental data: ATIC2 [57], JACEE [60], KASCADE [61], Tunka-133 [20], MSU [62], upper and lower limit for Auger's <lnA> is taken from [63].

The most uncertain point in our consideration is whether there are SNRs in our list of nearby sources that are able to accelerate protons up to 4 PeV. We included shell SNRs J1713-3946 and Vela Jr. (G266.2−1.2, RX J0852.0-4622) as accelerators with $E_{max}$=4 PeV based on theoretical predictions [64], [65]. These SNRs are not seen in Figure 4 because they are located at distances of around 1 kpc and their contributions to the observed all particle spectrum are less than 0.3 %. However, the distance and age of Vela Jr. is still under debate. This shell-type SNR (may be of Ib/c type) has a massive stellar progenitor and it is observed practically in all wavebands, including TeV gamma-rays with energies up to 30 TeV [42]. Firstly, $R$~ 250 pc and age ~ 680−1000 years were estimated in [66]. Later the ages (1.7−4.3) $10^3$ yr and distance ~750 pc, were derived that allowed the description of the details of filamentary structure and X-ray emission [67], [68]. There are theoretical estimates [64], where this source is located at $R$~1 kpc and $T$=3.34×$10^3$ yr. The possible contribution of Vela Jr. for two cases ($R$=0.3 kpc, $T$=0.7 kyr) and ($R$=0.7 kpc and $T$=1.7 kyr) are shown in Figure 4. One can see that, in the first case, Vela Jr. is the best candidate to be a knee producer; it can play the role of the 'single source', proposed in [24]. In contrast to ideas propounded in [24], the sharp shape of the knee in our model (see Figure 1) arises because particles with energy less than 1 PeV are still confined in this SNR if it is very close. The large contribution of Vela Jr. should be seen in anisotropy, which reaches then an extremely high value of tens of percent, as will be discussed below, see also Figure 7.

Theoretical predictions of PeV cosmic rays are absent for the closest shell SNRs with an age of 7−40 kyr: Cygnus loop (0.56 kpc); HB9 (0.8 kpc); HB21 (0.8 kpc); and S147 (0.8 kpc). The individual contribution of any of them is less than 2%; even if all four SNRs accelerate up to 4 PeV, they cannot significantly change the shape of the knee as it is seen in Figure 4. Note that they are too old to be detected in TeV gamma-rays.

In the bottom panel of Figure 4 we demonstrate (as an example) the energy dependence of <lnA> corresponding to the all particle spectrum presented in the top panel of Figure 4. To describe simultaneously the spectrum and <lnA> (E) we should make two important assumptions: the chemical composition before the knee (~1-3 PeV) should be light: protons plus helium nuclei should have the relative abundance ~65±5%, the abundance of iron nuclei should be ~ 10-16%, CNO and the middle group nuclei should be at the level ~ 15-20 %. Among the light nuclei, a dominance of helium nuclei (~ 18% of protons and ~ 48% of helium) before the knee is preferable. It provides more pronounced structure at around 20 PeV and better describes the energy dependence of <lnA>. The value of <lnA> reaches its maximum <lnA>~2.5±0.5 at $6×10^7$ – $10^8$ GeV due to the increasing contribution of galactic SNR IIb and protons of extragalactic origin. In the range $10^8$-$10^9$ GeV the <lnA> decreases, but a behavior of <ln A> above the energy $10^9$ GeV is mainly determined by the admixture of extragalactic cosmic rays (in Figure 4 we present two cases: without extragalactic protons and with extragalactic protons) and by the chemical composition of CR accelerated in SNR IIb. Here it can be emphasized that, due to the rarity of these explosions, the overwhelming contribution from a single source is expected [69]. We showed in [69] that Cas A is probably the best candidate to produce cosmic rays in the transition region since, of all young galactic SNRs, only Cas A is identified as a type IIb SNR [4], [5] (similar to SN1993J). Due to the complicated evolution of SNIIb progenitors, the chemical composition can differ from the average cosmic ray composition. In any case, the energies above $10^9$ GeV require special consideration, because of the probable significant contribution from extragalactic sources.

## 4. Anisotropy

### 4.1 Approach

Amplitude of a large-scale anisotropy in the approximation of isotropic diffusion is given by the equation:

$$\boldsymbol{A} = 3\frac{D}{c}\frac{\boldsymbol{\nabla} N(\boldsymbol{r},E)}{N(\boldsymbol{r},E)}$$

(see [1]).

[1]). The anisotropy depends on the gradient of CR density $\nabla N(\boldsymbol{r},E)$ and on the diffusion coefficient $D(E)$. In a course of their propagation from the sources to Earth, cosmic rays are scattered by random galactic magnetic fields. The particles are predominantly scattered by inhomogeneities with wave vectors $\boldsymbol{k}$ satisfying resonant condition $|\boldsymbol{k}|\sim 2\pi/r_g(E)$; here the gyroradius $r_g(E)=pc/eZB$ equals $\sim 2\ 10^{-3}$ pc at energy 10 TeV and 2 pc at 10 PeV; in the typical interstellar magnetic field $B=5$ μG, $p$ is the particle momentum. The diffusion mean free path $l = 3D/v$ ($v \approx c$ is the particle velocity) is equal to 24 pc at 10 TeV and 240 pc at 10 PeV for protons. Locally, CR diffusion is anisotropic since $r_g \ll l$ but strong random magnetic field with the principal scale $L \sim 100$ pc isotropies diffusion on larger scales $s>L$. In general, the diffusion approximation is valid at distances $s \gg l$ and times $t \gg l/v$ but in some cases these conditions are softer. In particular, the diffusion approximation is applied at small distances from an instant source $d<l$ if its age $t$ is larger than $l/v$ [70]. In our calculations, the last condition is not satisfied for 10 PeV protons only for sources closer than 250 pc at ages less than 800 yr, so the diffusion approximation can be applied even for Vela Jr. if it is at small distance $d = 300$ pc.

One of the sources of anisotropy is the Compton-Getting effect [71] when a dipole anisotropy is caused by the motion of the Earth through the isotropically distributed cosmic rays. The Earth's motion with velocity u~28 km/s around the Sun leads to the Solar Dipole (SD) anisotropy $\Delta I/I \sim \cos\theta(\gamma+2)u/c$, where $\theta$ is the angle between the cosmic ray arrival direction and the direction of motion; and $\gamma$ denotes the spectral index of CR flux. SD anisotropy could be observed in a reference system where the location of the Sun is fixed, e.g. where the latitude coordinate is the declination, and the longitude coordinate is defined as the difference between the right ascension of the cosmic ray arrival direction and the right ascension of the Sun, $\alpha-\alpha_{sun}$. In this reference frame, the excess is observed at 18 hr at the level $\sim 3.6\ 10^{-4}$, see e.g. [72]. This effect does not distort anisotropy in the sidereal time if the measurements are performed the whole year round [72]. The anisotropy that could be produced by the motion of the Solar System around the galactic center with a speed 220 km/s was not detected, e.g. [72]. This means that cosmic rays are involved in the general motion of the interstellar medium around the galactic center, through the scattering on magnetic fields frozen in the interstellar gas. The motion of the Sun with respect to the nearby stars and interstellar gas produces the Compton-Getting anisotropy of the order of $3\times 10^{-4}$ directed to $\sim 17-18$ hours in the sidereal time. This effect is taken into account in our calculations although it does not essentially influence the results.

After subtracting the Compton-Getting effects, the cosmic ray anisotropy is mostly determined by global leakage from the galaxy and by the contribution of nearby sources [1], [9], [10], [11]. Whether the first or the second contribution dominates depends on the specific source distribution in space and on the choice of the halo size, but in every case the very close and young sources make the energy dependence of anisotropy amplitude very irregular with bumps and dips. The specific new features of our present calculations are the work with an extensive list of potential cosmic ray sources compiled from a few catalogues of astronomical sources of non-thermal radiation, and the analysis of the procedure of cosmic-ray anisotropy measurements.

**4.2 Methods of measurements of anisotropy**

A large scale dipole anisotropy in the distribution of cosmic ray arrival directions on the sky of the order of $10^{-4} - 10^{-3}$ was traditionally measured in the past in the 'one-dimensional' observations as a sidereal time variation during the Earth rotation [73], [74], , [75], [76].In recent years, the series of advanced 'two–dimensional' anisotropy measurements have been performed in the northern sky in the Super-Kamiokande [12]; Tibet [13]; Milagro [14]; and ARGO-YBJ [15] experiments; and in the southern sky by the Ice Cube collaboration [16], [17], [77]. The anisotropies in these experiments have been measured in the TeV–PeV energy range both on large and small scales. The global anisotropies have an amplitude of the order of $10^{-3}$ and are smaller by a factor of a few to ten on scales between $10^o$ and $30^o$. The experimental procedure is based as a rule on measurements of the daily counting rate variations of the muons (or small EAS). As the Earth rotates, a fixed direction telescope in the horizontal coordinate system travels on the celestial sphere along the line of a constant declination, $\delta$, and returns to the same right ascension, $\alpha$, after one sidereal day [12]. A variation, $\varepsilon(\alpha, \delta)$, of the counting rate $n(\alpha, \delta)$ relative to the average counting rate at the fixed declination $<n(\delta)>$ is usually examined:

$$\varepsilon(\alpha, \delta) = \frac{n(\alpha,\delta) - <n(\delta)>}{<n(\delta)>} \qquad (11)$$

The anisotropy in the celestial sphere is studied in different experiments in one or two dimensions. The 'two-dimensional' map corresponds to $\varepsilon(\alpha, \delta)$ obtained for the fixed bands of $\delta$ with some steps $\Delta\delta=10^o$ in different intervals of $\delta$: $\delta= -50^o \div +90^o$ in Super Kamiokande [12]; $\delta=-15^o \div +65^o$ in Tibet [13]; $\delta= -80^o \div +80^o$ in Milagro [14]. The fist measurements in the southern sky performed by the Ice Cube collaboration were fulfilled recently in the interval of declinations $\delta= -80^o \div -30^o$ [14]. The term 'one-dimensional anisotropy' means that the measurements were performed in some wide range of declinations, as, for example, in the Baksan experiment [75], [76].

**4.3 Method of calculation of 'two-dimensional' anisotropy**

Using the Green function $G(t_i, \mathbf{r}_i, E)$ in the form (2) for the i–source with the time of cosmic ray emission $t_i$ and the coordinate $\mathbf{r}_i$ (the Solar System is located at the origin of the coordinates), one can calculate the $x, y, z$ anisotropy projections for all $N_{sour}$ sources as

$$A_x(E) = \sum_{i=1}^{Nsour} 3\frac{D}{c} \nabla_x \left( G(t_i, r_i, E) \right) / N(\mathbf{r}, E) \qquad (12)$$

and thus obtain the vector **A** that describes the amplitude and direction of anisotropy. The anisotropy was first calculated in the local *x, y, z* Galactic coordinate system and further transformed to the second equatorial coordinate system where the right ascension, $\alpha$, is expressed in hours and the declination, $\delta$, is expressed in degrees. This anisotropy will be called a '3-D' anisotropy.

At the next step we reproduced the experimental procedure of anisotropy measurements and obtained a '2-D' anisotropy in the following way. The grid of declinations $\delta_k$ with $\Delta\delta=10^o$ strips and the right ascensions $\alpha_j$ with $\Delta\alpha= 1$ hour strips was constructed and the projection of the i−source to the direction of the cell with coordinates $(\alpha_j, \delta_k)$ was calculated introducing the angle, $\vartheta_{ijk}$, between the direction toward the i-source and towards the $(\alpha_j, \delta_k)$ - cell:

$$\cos\vartheta_{ijk} = \sin\delta_i \sin\delta_k + \cos\delta_i \cos\delta_k \cos(\alpha_i - \alpha_j) \quad (13)$$

Then the amplitude of anisotropy toward the direction to the $(\alpha_j, \delta_k)$ - cell was calculated as

$$A_{k,j} = \frac{\sum_i 3D\frac{\partial}{\partial r_i}G(r_i,t_iE,)(\sin\delta_i \sin\delta_k+\cos\delta_i \cos\delta_k \cos(\alpha_i-\alpha_j))}{N(r,E)} \quad (14)$$

We then averaged $A_{k,j}(E, \alpha_j, \delta_k)$ over $\alpha_j$ in the bands with the fixed declination $\delta_k$ and obtained $<A_k(E, \delta_k)>$. The value of $\varepsilon(E, \alpha_j, \delta_k)$, defined similarly to the experimental variations, was calculated as

$$\varepsilon(E, \alpha_j\delta_k) = A_{k,j}(E, \alpha_j\delta_k) - <A_{k,j}(E, \delta_k)> \quad (15)$$

Due to the experimental procedure, the function $\varepsilon(E, \alpha_j, \delta_k)$ is the same for $\delta_k$ and $-\delta_k$.

We calculated the 'two-dimensional' map of the sidereal anisotropy based on Eq. (14) and (15) in the southern and northern hemispheres and added some random statistical fluctuations. The last are the spurious fluctuation, examined in [17], [77] by studying the dependence of counting rate on anti-sidereal time. A level of such fluctuations is of the order of $10^{-4}$ at the multi-TeV energies [17]. The 'two–dimensional' map at energy 10 TeV with the amplitude of random fluctuations $10^{-4}$ in small cells (0.5 hour $\times$ 5$^o$) is presented in Figure 5 in the top panel.

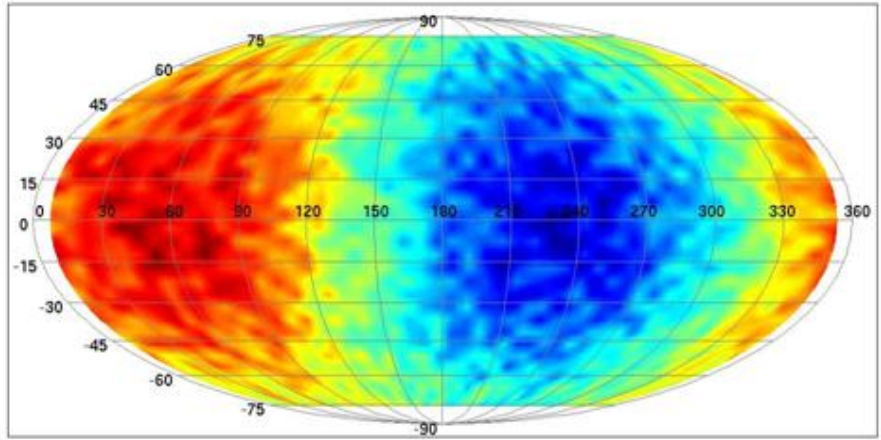

Fig. 5. [Top panel: '2D' anisotropy map at energy 10 TeV for the nearby sources listed in Table 1, with spurious fluctuations ~$10^{-4}$. The left spot marks the excess with amplitude A ~ $10^{-3}$ centered at $\alpha_j$ =3.2 hr (50$^o$), $\delta_k$ =0; the right spot marks the deficit with A~ -$10^{-3}$ centered at $\alpha_j$ =15.2 hr (230$^o$), $\delta_k$ =0.

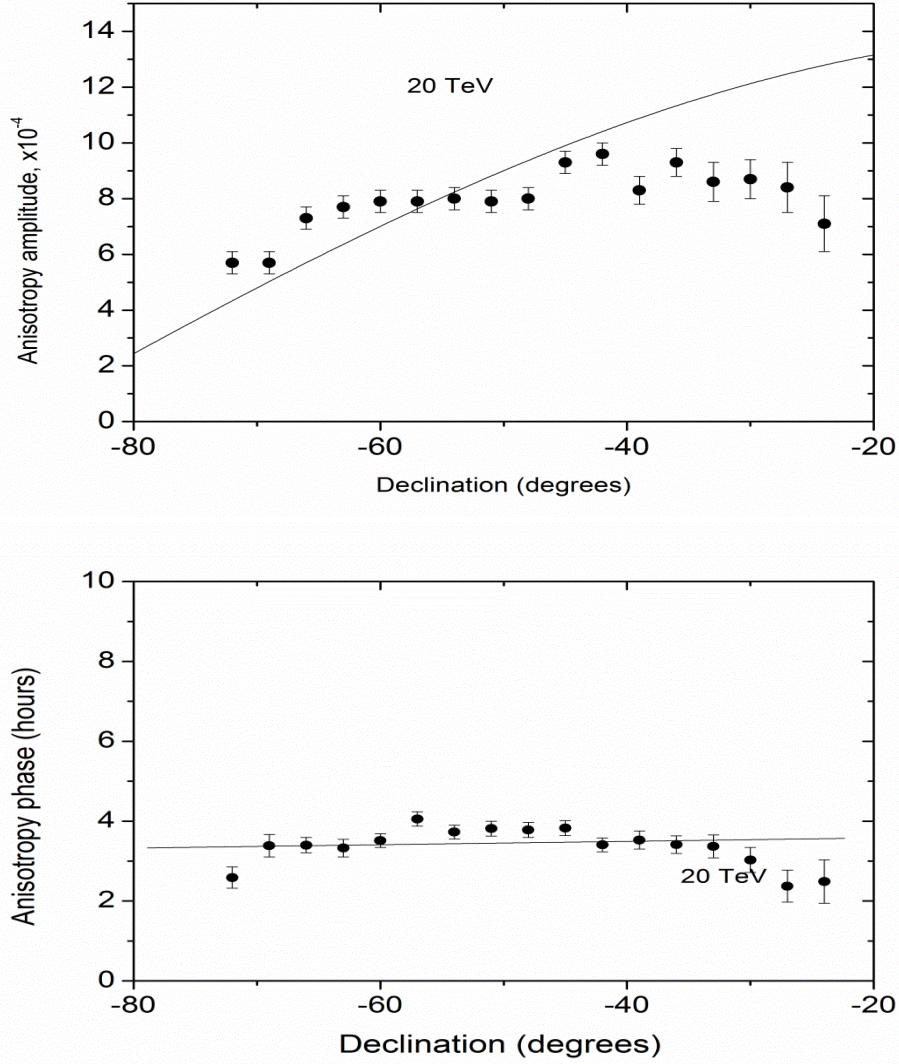

Fig. 5. [Middle panel: Anisotropy amplitude, $A(\delta)$, per declination band, obtained in Ice Cube [17] for the energy 20 TeV (black points); our calculation (lines) .[Bottom panel]: Anisotropy phase, $\varphi(\delta)$, per declination band [17] for $E \sim 20$ TeV (black points); our calculations (lines).

Then we fit the relative intensity, $\varepsilon(E,\alpha, \delta_k)$, in the declination band, $\delta_k$, by the usual harmonic function:

$$\varepsilon(E, \alpha, \delta_k) = A(E, \delta_k) \times \cos(\alpha - \varphi(\delta_k)) \qquad (16)$$

and obtain the amplitude, $A(\delta_k)$, and the phase, $\varphi(\delta_k)$, in every declination band. In accordance with Eq. (13-15) $\varepsilon(E,\alpha,\delta_k)$ represents a pure dipole component, the value of $A(\delta_k)$ is proportional to $cos(\delta_k)$, so the maximum of the amplitude is observed at the celestial equator, $\delta_k=0$; the phase of anisotropy does not depends on the declination, $\varphi(\delta_k)= const$. As a result the distribution of anisotropy amplitude resembles two wide spots. In Figure 5 (top panel) the left spot marks the relative CR excess with the maximum amplitude $\sim 10^{-3}$ centered at $\alpha_j$ =3.2 hr (50°), $\delta_k=0$; the right spot marks the deficit of CRs $\sim -10^{-3}$ centered at $\alpha_j =15.2$ hr (230°), $\delta_k=0$. The spurious fluctuations disturb a pure sinusoidal dependence but on the whole the calculated

'two-dimensional' map looks similar to the experimental ones as presented in different experiments [12] - [17]. This expectation seems to be confirmed by the new Ice Cube data [17]. The amplitude, $A(\delta_k)$, and the phase, $\varphi(\delta_k)$ of the first harmonic measured in Ice Cube are presented in the middle and bottom panels in Figure 5 together with our simulations, denoted by lines. Independence of the phase $\varphi(\delta_k)$ on $\delta$ is well seen. Coincidence is worse for the amplitude in the region of declinations $(-25 \div -35)^o$, which seems to be a boundary region for the effective anisotropy measurements. The authors of [17] emphasized that the region of declinations above $-25^o$ is masked due to the degradation of the angular and energy resolution at higher declinations. Such degradation is expected because of the poorer statistical power and the domination by mis-reconstructed events. Notice that only statistical error bars are presented in Figure 5.

It is necessary, however, to keep in mind that the comparison of CR counting rates between different declination bands is not fully justified. As stressed in [12], the 'two-dimensional' analysis is insensitive to the variation along the equatorial meridian; the 'two-dimensional' anisotropy must be considered as a series of one-dimensional curves in consecutive strips of declination. While the overburden rock above the muon detector in different declination belts varies, the median energy can also vary, sometimes very significantly – by an order of magnitude [12]. It is interesting to note that, in spite of this understanding, the data obtained in many experiments are interpreted 'directly', as if a true direction of anisotropy is measured (see the critical discussion of the Baksan group in [76]). For example, the two-dimensional anisotropy presented by Super Kamiokande [12] with the excess at ($\alpha = 75°\pm 7°$, $\delta = -5°\pm 9°$) is interpreted as a real direction of anisotropy to the Taurus Constellation [12] in the frames of the so-called NFJ model [79], proposed for the explanation of the "excess cone" and the "deficit cone" on the isotropic background measured in the muon experiments.

### 4.4. Anisotropy produced by ensemble of random sources

We start modelling of cosmic ray anisotropy energy dependence with the case when it is produced by the ensemble of point galactic sources with random positions and ages (as in [6], [7], [8], [10], [11]). 20 random realizations were simulated of sources with a birthrate of 1/50 yr$^{-1}$, a minimum distance of 100 pc from the Solar System and a minimum age of 50 years. A few important features were included in our calculations:

1) Experimental methods of anisotropy measurements were accounted for (including '2-D' and '3-D' procedures for each realization);
2) The arm structure of source distribution;
3) The energy dependent time of particle ejection from a source.

The compilation of experimental data is shown in Figure 6 for comparison with our calculations. Data on anisotropy amplitude and direction obtained in the underground experiments (UG) at energies from 300 GeV to 2 TeV are taken from the compilation [13]. Anisotropy measurements are shown above 2 TeV of the Super Kamiokande [12], Baksan [75], Argo-YBJ [15], Tibet As$\gamma$[13], EAS TOP [78], and IceCube [16], [17] experiments. The anisotropy around the knee obtained in different EAS experiments is taken from the compilation [80]. Two lines with symbols KASCADE and KASCADE Grande [80] represent the upper limits on the amplitude of anisotropy; the direction of anisotropy was not determined in these experiments. At PeV energies, the direction of anisotropy $\alpha \sim 230$ degrees ($-8.7$ hours) was determined only in the AKENO experiment [81], but this direction was obtained only for EAS with rich muon content (presumably produced by heavy nuclei). No statistically meaningful anisotropy was found for muon-poor showers (produced by the light components). It should be stressed that in two experiments, EAS TOP [78] and Ice Cube [77], at energy 100 – 300 TeV the direction of anisotropy changes rapidly: an excess at $\alpha \sim$ 1-3 hours jumps to an excess at $\alpha \sim -8 \div -12$ hours; at that time the absolute value of amplitude is very small at $4 \times 10^{-4}$. The Auger measurements [82] of large-scale anisotropy at very high energies are also shown.

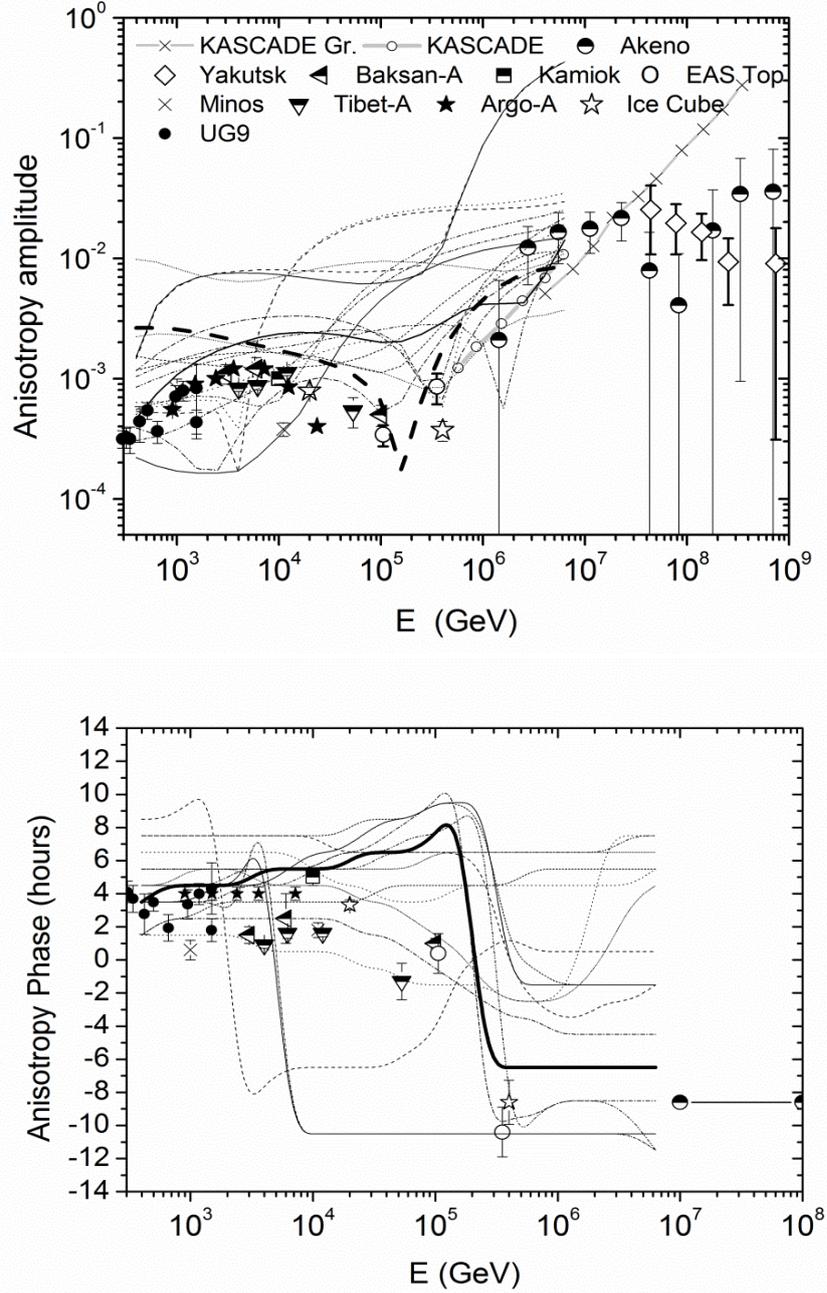

**Fig. 6.** Amplitude [top panel] and phase [bottom panel] of anisotropy produced by 20 realizations of point galactic sources with random positions and ages (one curve - one ensemble). Experimental data are denoted by symbols (see description in the text).

The results of our calculations of the amplitude, $A(E, \delta_k = 0)$, and the phase, $\varphi(E, \delta_k = 0)$, of anisotropy, at celestial equator for the primary protons are shown in Figure 6. Each curve representing the amplitude of anisotropy has an irregular energy dependence with bumps and dips caused by nearby sources. The direction phase, of anisotropy changes sharply for each curve. The '2-D' procedure of measurement reduces the real '3-D' amplitude by a factor of 1.5 - 2 times on average, because the contributions of sources located at high declinations, i.e. with large values of $|\delta_i|$, are suppressed in comparison with the sources located near the celestial equator. Besides, the measurements in declination belts located far from the equator, at $|\delta_k|>0$, reduce an amplitude by the additional factor $\cos \delta_k$. The inclusion of time dependent emissions in the calculations leads to the narrowing of bumps and even to the appearance of some narrow peaks in comparison with a case of instantly injected spectrum of all particles, since young and close sources do not have enough time to emit the low energy cosmic rays (see Figure 1 and discussion). In all experiments where only a median energy is reconstructed, the accuracy of energy reconstruction amounts to one order of magnitude [12], [13]; this procedure probably smoothes the experimental energy dependence.

A diversity of theoretical curves is of one order of magnitude at TeV energies and two orders in the PeV energy region. Analyzing the set of upper curves, we see that they correspond to the cases when one very young (<1 kyr) and close (<0.25 kpc) source appears near the Earth. The predominant direction (phase) of calculated anisotropy is $\alpha = 2 \div 6$ hours at energies from 100 GeV to $3 \times 10^4$ GeV in all 20 random samples. At higher energies, $3 \times 10^4$ GeV to $3 \times 10^6$ GeV, the direction of anisotropy jumps to $\alpha = 18$ hours in ten of eighteen samples. This change corresponds approximately to the drastic shift of anisotropy maximum from the Orion arm direction to the direction of the galactic center. The calculations with a uniform source distribution show that, in this case, the cosmic ray flux extends predominantly from the galactic center ($\alpha = 18 \pm 1$ hours) at all energies; this is in contradiction to observations. Besides, the amplitude of anisotropy becomes larger by a factor ~1.5 if we do not take into account the arm structure.

We also studied two additional factors: the influence of source density $q$ on anisotropy amplitude; and the influence of charge of primary particles. The results confirmed the analytical estimates made in [8]: the anisotropy amplitude increases approximately as $q^{-2}$. As regards the second question we have confirmed that the 'bump' produced by the nucleus with a charge Z is usually shifted to the higher energy by factor Z relatively the proton 'bump'; and at fixed energy the presence of nuclei diminishes the amplitude of anisotropy and smoothes the bumps.

**4.5. Anisotropy with a contribution of real nearby sources**

Let us now take into account information listed in Table 1 about the actual nearby sources with $R_{SNR} < 1.5$ kpc and $T_{SNR} < 6 \times 10^4$ years. In Figure 6 we presented anisotropy for primary protons only. The experimental data are obtained via the muon component up to $10-100$ TeV and by EAS above this energy. A muon spectrum is sensitive to the energy per nucleon, but not to the energy per primary nucleus so, practically, it is sensitive only to primary protons. Some data on anisotropy around 100 TeV (as in the Baksan [75], [76] and EAS TOP [78] experiments) are obtained with the small EAS, those also are mainly sensitive to primary protons. So the threshold region 100-500 TeV – is rather indefinite due to the unknown influence of nuclei on measured anisotropy. In the energy region above 500 - 1000 TeV, however, the chemical composition should be taken into account. To take this fact into consideration in the calculations presented in Figure 7, we assumed a proton dominance up to 100 TeV and a mixed composition of primary particles above 100 TeV, as it follows from our modeling (see Figure 4, bottom panel).

Ten curves illustrating the energy dependence of anisotropy are presented in Figure 7. Each curve corresponds to one set of sources that includes local sources from Table 1 and background sources simulated randomly. The amplitude of anisotropy (top panel) and the phase, (bottom panel) are shown (here the scale of right ascensions $0 \div 24$ hours was changed to the scale $-12 \div +12$ hours).

The calculated amplitude of anisotropy is of the order of $10^{-3}$ in the TeV energy range. Although our predictions fit the experimental values, the expected fluctuations remain of the order of $4\times 10^{-4}$. These fluctuations are due to the contribution of invisible sources with ages larger than $T_{near} \sim 6\times 10^4$ years and with large propagation times of the order of $10^5$ years at TeV energies (see Figure 2). Thus, the behavior of anisotropy amplitude with an accuracy better than $4\times 10^{-4}$ is impossible to predict.

The additional uncertainties are connected with the unknown parameters for the given remnant ($E_{max}$ value, the intensity, the $T_{esc}(E_{esc})$ dependence of accelerated cosmic rays) mainly determining the shape of 'bumps' produced by nearby sources, as it is seen in Figure 1 for the Vela X remnant. In the bottom panel of Fig. 7 it can be seen that, in some cases of calculations, there are narrow bumps located at right ascension ~ 8.5 hours, caused by Vela X, if it accelerates particles up to $E_{max} \sim 100$ TeV with intensity less than an average SNR by tree times, and if we assume $T_{esc}(E_{esc})$ as (8). In (8) particles with $E<10$ TeV are confined in the Vela X at age 11 kyr. Consequently, gamma - ray telescopes should detect gamma - ray emission not higher than 1 TeV. The telescope HESS detected gamma-ray emissions from this remnant in the energy range 0.75÷20 TeV from the central region only [83], and this should surely be interpreted as a gamma-ray emission produced by the population of high energy electrons in PWN [83], and not by nuclear components at the outer shock front. Some authors relate the Vela X to the class of Crab-like events [51]. The fact that we do not see the wide bump in anisotropy from Vela X *($\alpha$~8.5 hours)* in some sense confirms our suggestions that the diversity of SNR properties should lead to a diversity of emitted spectra of cosmic rays from different remnants.

To put some limits for this diversity we suggested that all composite SNRs and PWNs (as well as nearby sources of this type: Boomerang, CTA1, IC443, Vela X, J1708-443), can accelerate CR with a power reduced by 2–3 times in comparison with the shell SNRs. This reduction does not change noticeable the diversity of predictions presented in Fig. 7, but if we would suppress the contribution of composite SNRs and PWNs by 10 times, then neither the amplitude nor the phase of the observed anisotropy can be reproduced. In this case the CR gradient for all energies is directed from the center of the galaxy and the amplitude of anisotropy exceeds the measured amplitude by 2–3 times. It means that, not only SNR of the Ia type provide total CR intensity, but also SNR of type II are efficient CR emitters.

The experimental phase of anisotropy is around $\alpha \sim 2 \div 6$ hours (see bottom panel of Figure 7). It is caused by the bulk of nearest sources located in the 'Orion arm' (see Figure 3 and the discussion in chapter 4.4); their coordinates correspond to the equatorial coordinates $\alpha \sim 0 \div 4$ hours and $\delta \sim 60 \pm 5^o$ for sources located in the galactic disk. In '2D' approach it is impossible to determine the value of declination $\delta$. But in the Baksan experiment [84], just the same value of $\delta$ was obtained experimentally, where both coordinates of the direction (right ascension and declination) were estimated by the special method based on the analysis of a zero harmonic of intensity in the sidereal time. It was found that the direction of anisotropy vector approximately corresponds to $\delta \sim 60^0$ and $\alpha \sim 3$ hours. This proves that the cosmic ray anisotropy at multi-TeV energies is caused by the disk population of sources in the Orion arm.

At higher energies of 10 TeV to 500 PeV, the modelling reveals the dip of anisotropy amplitude and the corresponding change of excessive flux direction from the Orion arm to the center of the galaxy in many random realizations of cosmic ray sources. Such kind of behavior was seen in two experiments, EAS TOP [78] and Ice Cube [17], [77], where the rapid changes in the direction of anisotropy occurred at energies 100 to 300 TeV. If we assume a uniform spatial distribution of sources ('No arms' in Figure 3), the anisotropic flux dominates from the center of the galaxy ($\alpha$=18 ±1 hours) practically for all samples and at all energies that contradicts the experimental data.

The typical propagation time from the source to the observer is around $10^4$ years at PeV energies; the fluctuations caused by old invisible sources are relatively small. The level of fluctuations at these energies is determined in our calculations by the unknown values of $E_{max}$ in nearby sources.

Because the exact types of the nearby SNRs described in Table 1 with corresponding $E_{max}$ are unknown, we first fixed $E_{max}$ randomly, assuming that the ratio of SNRIa to other SNRs is the same as for all populations of distant sources. Figure 7 shows 10 random realizations of sources, when one nearby source accelerates particles to $E_{max} = 4 \times Z$ PeV. On average, the predicted amplitude is in the range 0.5% to 5 % and approximately reproduces observations. The direction of anisotropy in this case is determined by this specific source, the name of which is written near the curve in Figure 7. If one of the sources R5 ($\alpha=1.5$ h); HB9 ($\alpha =5$ h); S147 ($\alpha =5.7$ h); Vela Jr. ($\alpha =8.6$ h); or IC443 ($\alpha =6.3$ h) accelerates up to $4 \times Z$ PeV, the direction of anisotropy contradicts the AKENO data. The AKENO data can be fitted if the Cygnus loop ($\alpha =-4$ hr), J1713 -3949 ($\alpha =-7$ hr) or HB21 ($\alpha =-4$ hr) accelerates up to $4 \times Z$ PeV.

Using another method of calculation, we considered two extreme cases: 1) absence of Ia type SNR with $E_{max}= 4 \times Z$ PeV among the nearby sources in the radius R $\leq 0.8$ kpc; 2) all shell SNRs Cygnus loop, HB9, HB21, S147 and Vela Jr. accelerate nuclei up to $4 \times Z$ PeV. The expected number of SN Ia at $R \leq 0.8$ kpc and with $T< 6 \times 10^4$ yr is ~ 1 in our model (for total SN birthrate 1/50 yr$^{-1}$ and 20% of SNRIa among them), so the second assumption is more probable than the second. Two extreme cases give more or less similar results regarding the amplitude of anisotropy and do not contradict the experimental values. Unfortunately, the measurements of the direction of anisotropy around the knee are practically absent.

The diversity of theoretical predictions for anisotropy around the knee cannot be reduced, since we do not know $E_{max}$ for every actual source but, in spite of the differences seen among different cases described above, there are some common features. The calculated amplitude of anisotropy, (0.5- 5)% at energies $10^6$-$10^8$ GeV, does not contradict the experimental values.

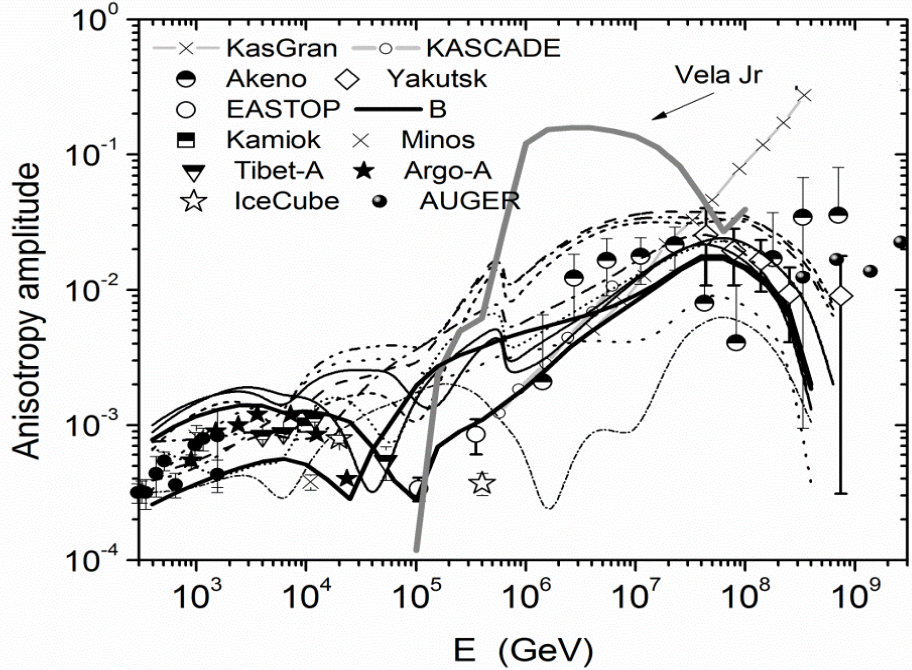

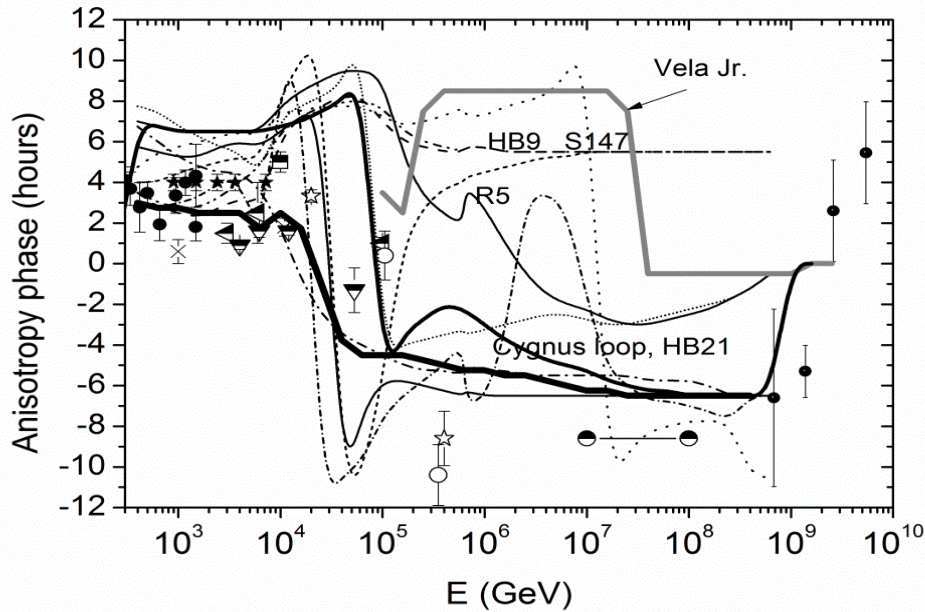

**Fig. 7.** Amplitude [top panel] and phase [bottom panel] of cosmic ray anisotropy. Experimental data (symbols, see description in the text); dipole anisotropy calculated at celestial equator ($\delta_k=0$) for three extreme cases: 1) absence of sources with $E_{max}= 4\times Z$ PeV among nearby sources in radius $R<$ 0.8 kpc (very thick line); 2) all nearest shell SNRs: Cygnus loop, HB9, HB21, S147 accelerate CR up to $4\times Z$ PeV (thick line); 3) the case when Vela Jr. is located at $R=0.3$ kpc, has an age $T=0.7$ kyr and accelerates particles to $4\times Z$ PeV (gray thick line). Dashed and dotted lines represent the cases when one of nearby sources accelerates CR up to $4\times Z$ PeV (its name is written near every curve).

A relatively small value of anisotropy in our calculations is reached due to the following assumptions: 1) slow energy dependence of the diffusion coefficient $D\sim E^{0.33}$; 2) admixture of heavy nuclei around the knee; 3) absence of young nearby cosmic ray sources in the near vicinity of 0.5 kpc around the Sun; 4) relatively large number of SNR being able to accelerate particles up to 4 PeV; its rate is 1/250 yr$^{-1}$ (20% of SNR Ia among all SNR with a total SNR birthrate of 1/50 yr$^{-1}$). All these assumptions are in accord with the modern scenarios of cosmic ray origin.

The account of characteristics of actual nearby sources allows a significant reduction in uncertainty in the prediction of anisotropy. We failed, however, to find the set of nearby sources that can reproduce the dip with amplitude $\sim 3\times10^{-4}$ in the observed anisotropy at 100 to 300 TeV. It seems that efforts to fit axactly with the experimental data have no meaning, since many old background sources with ages more than $6\times 10^4$ years contribute to anisotropy at these energies. Besides, a position of the peak from a given source strongly depends on a poorly known $T_{esc}(E)$ dependence. The situation is better at PeV energies, where the propagation time is less than the life of a shell. More definite conclusions will be reached when the direction of anisotropy is accurately measured in the knee region.

It can be inferred from the above considerations that experimentally measured large-scale dipole anisotropy (both the amplitude and the direction) in the multi-TeV energy region is reproduced successfully in a simple diffusion model, when the main part of large-scale anisotropy is determined by the gradient of cosmic ray density caused by the global CR flow from the direction of the galactic center, supplemented with contributions of the nearest sources located in the Orion arm. It is important to note that in practically all experiments at multi-TeV energies, the presence of the second quadruple harmonic was found [12] - [17] at the level (2–4) $10^{-4}$ with the phase ~ 19 hours. This implies the presence of a small bipolar excessive flux from two directions: 7 hours and 19 hours. These directions coincide with the points where δ=0, that is connected with the excessive fluxes along the galactic plane. The second angular harmonics of cosmic ray anisotropy may arise in a course of cosmic ray diffusion through magnetic traps formed by a large-scale random magnetic field in the galactic disc [85].

## 5. Discussion and conclusions

The goal of this paper was to explain simultaneously the new cosmic ray data on a fine structure of all particle spectra around the knee [19], [20], [21], [22] and anisotropy in the wide energy range 1 TeV÷1 EeV. We develop the model of galactic cosmic ray sources, supernovae remnants, based on the main ideas proposed in [28]. The essential new feature of the model is non-fixed $E_{max}$ for cosmic rays accelerated in core collapsed supernovae SNII and SNIbc in the range 1 TeV ÷ 4000 TeV with a decreasing number of core collapsed SNRs being able to accelerate CR up to given energy E, that leads to a steeper total CR source spectrum ($\gamma$ ~ -2.4) in comparison with the spectrum of the individual source ($\gamma_{sour}$ ~ -2.2). The last value does not contradict the instant gamma-rays spectrum measured in the Tycho's SNR, but a simple non-linear diffusive shock acceleration theory predicts $\gamma_{sour}$ ~ -2.0 and the theory requires complication [27, 86, 87] to explain more steep source spectra.

The group of SNRs Ia constituting ~20% of all SNRs provides the spectrum between $4 \cdot 10^{15}$ eV and $10^{17}$ eV as well as a high content of Fe nuclei around $10^{17}$ eV. If we introduce a sharp cutoff of the source spectrum above $E_{max}$ ~ 4×Z PeV by $d\gamma$ ~ 2, it is possible to reproduce the fine structure of all particle spectrum, measured in the Tunka133 and KASCADE-Grande experiments.

The role of nearby galactic sources in the formation of observed energy spectrum and anisotropy is taken into consideration. The list of these sources is made up based on radio, X-ray and gamma-ray catalogues. The number of selected sources does not contradict the SNe birthrate in the galaxy 1/50 yr$^{-1}$. The distant sources are treated statistically as an ensemble of sources with random positions and ages. In comparison with previous similar works [9], [24], we took into account the experimental procedure of 'two-dimensional' anisotropy measurements, and analyzed not only amplitude, but also the direction of anisotropy, and included in the consideration the time dependent emission of cosmic rays from supernova remnants, which were very important for this task. We also took into account the arm structure of the galaxy that is also very important for the investigation of the anisotropy direction.

It is shown that only one source - Vela Jr., if it is very young and close (R~0.3-0.5 kpc, T~0.7-1.7 ky) is suited to the role of the 'single source' determining the 'sharpness' of the knee. In this case the sharp form of the knee arises because particles with energy less than 1 PeV are still confined in the SNR. But anisotropy reaches a very high value (tens of percent) and contradicts the small anisotropy measured around the knee.

The amplitude and direction of the dipole component of anisotropy, basically, is caused by nearby sources. We reproduced the main features of energy dependence of measured anisotropy before the knee: a small value at the multi-TeV range; the influence of the Orion arm; and the change of direction in the range 10-300 TeV. The applied model reproduced the anisotropy around the knee at energy $10^6$-$10^8$ GeV: the calculated amplitude of anisotropy, 0.5- 5%, does not contradict experimental values. A relatively small value of anisotropy amplitude can be obtained due to four important factors: 1) a slow energy dependence of the diffusion coefficient $D_{dif}$ ~ $E^{0.33}$; 2) the admixture of heavy nuclei around the knee; 3) absence of young nearby cosmic ray sources in close (<0.5 kpc) vicinity to the Sun; 4) the number of SNR being able to accelerate particles up to 4 PeV is rather large 1/250 yr$^{-1}$ (20% of SNR Ia amongst all SNR with a total SNR birthrate of 1/50 yr$^{-1}$). All these assumptions do not look exotic, and they confirm the modern understanding of cosmic ray origin.

In comparison with the recent studies of cosmic ray flux and anisotropy based on the numerical simulations of cosmic ray production in random SN explosions [10], [11], we take into account in the present work the actual nearby sources known from the latest radio, X-ray and gamma-ray catalogues. The latest results of the gamma-ray astronomy most likely exclude the presence of unknown SN remnants in the 1.5 kpc vicinity of the Sun. In our work we specify the most important sources (Vela Jr, Vela X, Cas A) that may contribute significantly to the spectrum and anisotropy of observed cosmic rays. The decreasing number of sources accelerating cosmic rays up to given energy was introduced in the present calculations. The fine structure of the all particle spectrum and the general features of the energy dependence of the dipole component of anisotropy are explained in our model with an accuracy of about a factor of 3.

It should be stressed that the isotropic diffusion assumed in our calculations is a serious simplifying assumption which is difficult to avoid because of a complicated and, in fact, unknown detailed structure of the galactic magnetic field. The presence of a large-scale random magnetic field justifies the approximation of isotropic diffusion at scales larger than a few hundred parsecs, but the anisotropy is a characteristic which is very sensitive to the local surroundings of the Solar System, including the orientation of the diffusion tensor and the values of its components.

**Acknowledgements.** The authors are grateful to the referee for very useful remarks, that helped us to considerably improve the article. This work was supported by RFBR grants 10-02-01443-a, 13-02-00056, 11-02-91332-NNIOa, 11-02-12138-ofi-m-2011, 13-02-12095-офи-м, and Russian Ministry of Education and Science Contract, № 14.518.11.7046.

**Appendix A. Green function in a simple galactic diffusion model.**

The basic model for the investigation of cosmic-ray propagation in the Galaxy is the flat halo diffusion model [1]. We consider here its simplest version, see Fig. A1. All cosmic ray sources lie in the infinitely thin galactic disk. The cosmic ray diffusion coefficient $D$ is constant in the entire Galaxy with a cosmic ray halo of the height $H$. Cosmic rays freely escape through the halo boundaries to the intergalactic space where their density is negligible. This implies the boundary condition $N|_\Sigma = 0$. Considering the case of a large galactic radius $R \gg H$, we set $R = \infty$.

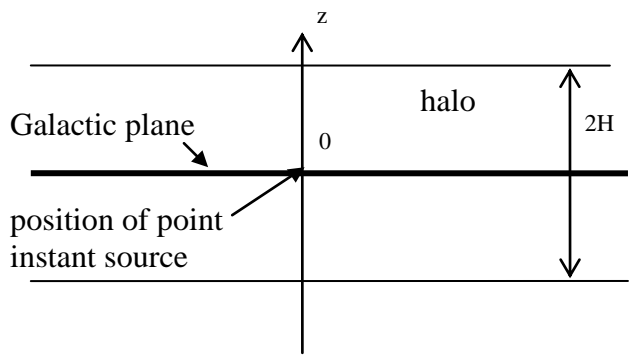

Fig. A1. Geometry of simple galactic diffusion model.

Let us find the solution $G$, the Green function, of diffusion equation (1) for a point instant source located at $x = 0$, $y = 0$, $z = 0$ (the axis $z$ is perpendicular to the galactic plane, the axes $x$ and $y$ are in the galactic plane). The equation takes the form

$$\frac{\partial G}{\partial t} - D\frac{\partial^2 G}{\partial z^2} - D\frac{\partial^2 G}{\partial x^2} D\frac{\partial^2 G}{\partial y^2} = \delta(z)\delta(x)\delta(y)\delta(t) \qquad (A1)$$

We search for a solution in the form of a series which automatically satisfies the boundary conditions at $z = \pm H$:

$$G = \sum_{n=1}^{\infty} F_n(t, x, y)\cos\left(\left(n - \frac{1}{2}\right)\frac{\pi z}{H}\right) \qquad (A2)$$

Taking into account that $\delta(z) = \frac{1}{H}\sum_{n=1}^{\infty}\cos\left(\left(n - \frac{1}{2}\right)\frac{\pi Z}{H}\right)$, one has now the following equation for $F_n$:

$$\frac{1}{D}\frac{\partial F_n}{\partial t} + \left(\left(n - \frac{1}{2}\right)\frac{\pi}{H}\right)^2 F_n - \frac{\partial^2 F_n}{\partial x^2} - \frac{\partial^2 F_n}{\partial y^2} = \frac{1}{DH}\delta(x)\delta(y)\delta(t) \qquad (A3)$$

.

The substitution $W_n = F_n \exp\left\{\left(\left(n - \frac{1}{2}\right)\frac{\pi}{H}\right)^2 Dt\right\}$ leads to the diffusion equation for $W_n$ that has the solution $W_n = \frac{\vartheta(t)}{4\pi DHt}\exp\left\{-\frac{x^2+y^2}{4Dt}\right\}$ (here $\vartheta(t)$ is the step function) and gives the desired Green function:

$$G = \sum_{n=1}^{\infty}\frac{\vartheta(t)\cos\left(\left(n-\frac{1}{2}\right)\frac{\pi Z}{H}\right)}{4\pi DHt}\exp\left\{-\left(\frac{x^2+y^2}{4Dt} + \left(\left(n-\frac{1}{2}\right)\frac{\pi}{H}\right)^2 Dt\right)\right\} \qquad (A4)$$

**References**

[1]. V.S. Berezinskii, S.V. Bulanov, V.A. Dogel, V.L. Ginzburg, V.S. Ptuskin, Astrophysics of Cosmic Rays, North-Holland, Amsterdam, 1990.
[2]. F. Aharonian. Very High Energy Cosmic Gamma Radiation: A Crucial Window on the Extreme Universe, World Scientific, 2004.
[3]. Mathieu de Naurois, for the H.E.S.S. collaboration, AIP Conf. Proc. 1223 (2011) 12-22.
[4]. J. Vink., Adv. Space Rec. 33 (2004) 356
[5]. J. Vink, Astronomy and Astrophysics Review 20 (2012) N 49
[6]. M.A. Lee, ApJ 229 (1979), 424–431
[7]. A.D. Erlykin, A.W.Wolfendale, J. Phys. G: Nucl. Part. Phys., 32, (2006), 1;
[8]. A. A. Lagutin, Y.A. Nikulin, JETP (1995) 825.
[9]. V.S. Ptuskin, F.C. Jones, E.S. Seo, R. Sina, Adv. in Sp. Research 37 (2006)1909.
[10]. P. Blasi and E. Amato, JCAP 01(2012) 010 and 011.; arXiv:1105.4529v1 [astro-ph.HE].
[11]. M. Pohl, D. Eichler, ApJ, 766, (2013), 4.
[12]. G. Guillian et al. (Super-Kamiokande Collaboration), Phys. Rev. D, 75 (2007), 062003.
[13]. M. Amenomori et al. (Tibet ASγ Collaboration), Science, 314 2006, 439-443; Astrophys. J., 2005, 626: L29–L32; astro-ph/ 0505114v1. 2005.
[14]. A. A. Abdo *et al.*, Astrophys. J. 698 (2009) 2121-2130.
[15]. R. Iuppa, G. Di Sciascio (ARGO-YBJ collaboration), Nucl. Inst. and Meth. in Physics Research 692, (2012)160.
[16]. S. Toscano and Ice Cube Collaboration, Nucl. Phys. Proc. Suppl. 212-213, (2011) 201-206.

[17]. R. Abbasi, Y. Abdou, T. Abu-Zayyad et all (Ice Cube collaboration), Astroph. J. 746 (2012) 33.
[18]. G. Giacinti and G. Sigl1. Phys.Rev.Lett. 109 (2012) 071101.
[19]. S.F. Berezhnev, D. Besson, N.M. Budnev et al (Tunka 133 collaboration), Proc. of 32 Int. Cosmic Ray Conf, Beijing, 2011, (Beijung:scientific) vol. 7, p. 208-211; p.197-200.
[20]. S.F. Berezhnev, D. Besson, N.M. Budnev et al (Tunka 133 collaboration); Nuclear Instruments and Methods in Physics Research A 692 (2012) 98 – 105.
[21]. W.D. Apel, J.C. Arteaga-Vel´azquez et al (KASCADE Gr. Collaboration), Astrop. Phys. 36 (2012) 183-194.
[22]. W.D. Apel, J.C. Arteaga-Vel´azquez et al (KASCADE Gr. Collaboration), Phys. ReV. Let. 107 (2011) 171104.
[23]. A. Garyaka, R. Martirosov, S. Ter-Antonyan et al, (GAMMA collaboration), Journal of Physics: Conference Series, 409, 1 (2013)012081.
[24]. A.D. Erlykin, A.W. Wolfendale, J. Phys. G: Nucl. Part. Phys. 23 (1997) 979.
[25]. A.D. Erlykin, A.W. Wolfendale, R. Martirosov R CERN-Courier Jan/Feb 201 (2011) 21.
[26]. V.S. Ptuskin, I.V. Moskalenko, F.C. Jones, A.W. Strong, V.N. Zirakashvily, Astrophys. J.642 (2006) 902-916.
[27]. D. Caprioli, JCAP, 07 (2012) 38.
[28]. V. Ptuskin, V. Zirakashvili, E.S.Seo. The Astroph. Journal, 718 (2010) 31.
[29]. D. Caprioli, P. Blasi, E. Amato, "On the escape of particles from cosmic ray modified shocks" MNRAS, 396, (2009) 2065.
[30]. V.N. Zirakashvili and V.S. Ptuskin, Astropart. Phys. 39-40 (2012) 12-21.
[31]. A.R. Bell, S.G. Lucek, MNRAS, 321 (2001) 433.
[32]. V.S. Ptuskin & V.N. Zirakashvili, A&A, 403 (2003) 1.
[33]. V.S. Ptuskin & V.N. Zirakashvili, A&A, 429 (2005) 755.
[34]. R. Chevalier, Astrop. J, 619 (2005) 839.
[35]. W. Li, R. Chornock, J. Leaman, A.V. Filippenko et al, MNRAS, 412 N3 (2011) 1473.
[36]. M. Hamuy, 'Core Collapse of Massive Stars', ed. by C.L. Fruer (Kluwer academic Publishers, Dordrecht) (2003).
[37]. R.J. McMillani et al. Astroph. J. 473 (1996)707.
[38]. D. A. Green. Bull. Astr. Soc. of India, 37 No. 1 (2009) 45-61, arXiv:0905.3699.
[39]. G. Case and M. Bhattacharya. A&A Supplement. Ser.120 (1996.) 437.
[40]. R. N. Manchester, G. B. Hobbs, A. Teoh and M. Hobbs, Astron. J. 129 (2005) 1993.
[41]. A. A. Abdo et al (Fermi collaboration), ApJ 187 (2010) 460.,
[42]. O. Kargaltsev and G. G. Pavlov, AIP Conf. Proc. 983 (2008) 171-185; .
[43]. H.E.S.S. home page. http://www.mpi-hd.mpg.de/hfm/HESS/pages/home/sources. F.Aharonian (HESS Collaboration) Astroparticle Physics 34 (2011) 738-747.
[44]. J.Albert et al (MAGIC col), Astrophys. J. 639 (2006) 761-765; arXiv:0508543.
[45]. V. Acciari (VERITAS col.) Ap. J Letters 730 (2011) L20.
[46]. N. Vranesevic, R. N. Manchester, D. R. Lorimer, Astroph. J., 617 (2004) L139–L142.
[47]. D. Poznanski, N.Bulter, A.V. Filippenko et al, Astrophys.J. 694 (2009) 1067-1079.
[48]. R. J. Mcmillan1 and R. Ciardullo, Astroph. journal, 473 (1996) 707È712
[49]. P. Englmaier, M. Pohl, and N. Bissantz, Memorie della Societa Astronomica Italiana Supplement, 18 (2011) 199.
[50]. http://astronet.ru/db/msg/1228169/MWspitzer_lab_2048.jpg.html
[51]. B. M.Gaensler and P.O. Slane, Ann. Rev. Astron. Astrophys. 44 (2006) 17-47.
[52]. G. Ferrand, and S. Safi-Harb, Advances in Space Research, 49, (2012). 1313-1319.

[53]. Samar Safi-Harb, AIP Conf. Proc. 1505 (2012) 13-20; arXiv:1210.5406 [astro-ph.GA]
[54]. N. Smith. arXiv:1304.0689v1 [astro-ph.HE].
[55]. J.J. Hester, ARA&A, 46, (2008), 127
[56]. F. D. Seward, P. Gorenstein & R.K. Smith, ApJ, 636 (2006) 873.
[57]. A. D Panov, J.H. Adams Jr., H.S. Ahn et al (ATIC2 Col). Bull. of the Rus. Academy of Sciences: Physics. 73 5 (2009) 564–567.
[58]. M. Amenomori, X. J. Bi, D. Chen et al (Tibet Col.) Ap J 678 (2008) 1165-1179.
[59]. E.E. Korosteleva et al ((Tunka 25 col.) Nuclear Physics B (Proc. Supp.) 165 (2007) 74-80.
[60]. Takahashi, Y. (JACEE Collaboration), Nucl. Phys. B (Proc. Suppl.), 60 (1998) 83; Asakimori, K., et al. 1998, ApJ, 502, 278.
[61]. Roth, V. and Ulrich, H., Proc. 28th ICRC, Japan, Tsukuba, 2003, vol. 1, p. 139.
[62]. N.N. Kalmykov, G.V.Kulikov, V.P.Sulakov et al. Moscow University Physics Bulletin, 64 N 5 (2009) 536-540.
[63]. P.Abreu and Auger collaboration. arXiv:1301.663.
[64]. E.G. Berezhko, G. P¨uhlhofer and H.J. V¨olk, A&A 505 (2009) 641-654.
[65]. V. N. Zirakashvili & F.A. Aharonian (2010), Astroph. J. 708(2) (2009) 965-980.
[66]. A. N. Cha, K. R. Sembach & A.C. Danks ApJ, 515 (1999) L25.
[67]. Th. G. Pannuti, G. E. Allen et al. Astroph. J. 721 (2010) 1492.
[68]. J. Better. S. Katsuda, H.Tsunemi, Mori K, Advances in Space Research, 43 (2009)895–899; arXiv: 2009.01.004.
[69]. L.G. Sveshnikova, E.E. Korosteleva, L.A. Kuzmichev, V.S. Ptuskin, V.A. Prosin and O.N. Strelnikova, Journal of Physics: Conference Series, 409, 1 (2013)012062.
[70]. Yu.I. Fedorov, B.A. Shakhov, Astronomy & Astrophysics, 402 (2003) 805
[71]. A.H. Compton & I.A. Getting, Phys. Rev., 47 (1935) 817.
[72]. R. Abbasi et al (Ice Cube col.), Proc. Of 32th ICRC, Beijing, 2011, N 305.
[73]. D. J. Cutler & D. E. Groom, Nature, 322 (1986) 434.
[74]. K. Nagashima et al., Nuovo Cimento C, 12 (1989) 695.
[75]. V.A. Kozyarivsky, A.S. Lidvansky, V.B. Petkov and et al., Proc. 29 Int. Cosmic Ray Conf. 2004, v. 2, p.93-96; arXiv:0406059.
[76]. V.A. Kozyarivsky, A.S. Lidvansky, Astronomy Astronomy Letters, 34, N2 (2008) 113-117. D.D. Dzhappuev, V.A. Kozyarivsky, A.U. Kudzhaev, A. S. Lidvansky, and T. I. Tulupova, Astronomy Letters, 36, N 6 (2010) 416-421.
[77]. R. Abbasi et al. (Ice Cube col.), Proc. of 32th ICRC, Beijing, 2011, N 305.
[78]. M. Aglietta et al., Astrophys. J., 692 (2009) L130; Journal of Physics: Conference Series 203 (2010) 012126.
[79]. K. Nagashima, K. Fujimoto, and R. M. Jacklyn, J. Geophys. Res. No. A8 103 (1998) 17429
[80]. S. Overe, M. St¨umpertc, W.D. Apela et al, Proc. of the 30th ICRC, Mexico, 2008, vol. 4 , p. 223–226.
[81]. T. Kifune, T. Hara, Y. Hatano et al , J. Phys. G: Nucl. Phys. 12 (1986) 129.
[82]. M. Grigat1 et al (Pierre Auger Collaboration). ASTRA. Astrophys. Space Sci. Trans., 7, (2011) 125–129.
[83]. H.E.S.S. Collaboration, A. Abramowski1 et al; arXiv:1210.1359.
[84]. V. A. Kozyarivsky, A. S. Lidvansky, T. I. Tulupova, Proc. Of 32th ICRC, Beijing (2011) N 281.
[85]. E.G.Klepach, V.S.Ptuskin, Astronomy Letters 21 (1995) 411-417.
[86]. G. Morlino and D. Caprioli, A&A, 538 (2012) 81;
[87]. E. Berezhko, Ksenofontov, Volk, ApJ, 763 (2013) 14)